\documentclass[journal]{IEEEtran}
\usepackage{amsmath,graphicx}
\usepackage[T1]{fontenc}
\usepackage{hyperref, times, cite}
\usepackage{url}
\usepackage{enumitem,kantlipsum}
\usepackage{mathtools}
\usepackage{bm, bbm}
\usepackage{cool}
\usepackage{amssymb,amsfonts,amsmath,amsthm,amscd,dsfont,mathrsfs}
\usepackage{graphicx,float,psfrag,epsfig,amssymb}
\usepackage{wrapfig}
\usepackage{relsize}
\usepackage{color}
\usepackage{pict2e}
\usepackage{algorithm}
\usepackage{algorithmic}
\usepackage{caption}
\usepackage{nameref}
\usepackage{enumerate}
\usepackage{stackrel}

\usepackage{caption}
\usepackage{subcaption}

\usepackage{amsthm}

\begin{document}
\thispagestyle{empty}

\title{On Sparse High-Dimensional Graphical Model Learning For Dependent Time Series}
\author{ Jitendra K.\ Tugnait 
\thanks{J.K.\ Tugnait is with the Department of 
Electrical \& Computer Engineering,
200 Broun Hall, Auburn University, Auburn, AL 36849, USA. 
Email: tugnajk@auburn.edu . }

\thanks{This work was supported by the National Science Foundation under Grants CCF-1617610 and ECCS-2040536.}}

\maketitle

\renewcommand{\algorithmicrequire}{\textbf{Input:}}
\renewcommand{\algorithmicensure}{\textbf{Output:}}

\begin{abstract}
We consider the problem of inferring the conditional independence graph (CIG) of a sparse, high-dimensional stationary multivariate Gaussian time series. A sparse-group lasso-based frequency-domain formulation of the problem based on frequency-domain sufficient statistic for the observed time series is presented. We investigate an alternating direction method of multipliers (ADMM) approach for optimization of the sparse-group lasso penalized log-likelihood. We provide sufficient conditions for convergence in the Frobenius norm of the inverse PSD estimators to the true value, jointly across all frequencies, where the number of frequencies are allowed to increase with sample size. This result also yields a rate of convergence. We also empirically investigate selection of the tuning parameters based on the Bayesian information criterion, and illustrate our approach using numerical examples utilizing both synthetic and real data. 
\end{abstract}


\section{Introduction}
Graphical models are a useful tool for analyzing multivariate data \cite{Whittaker1990}, \cite{Lauritzen1996}, \cite{Buhlmann2011}. A central concept is that of conditional independence. Given a collection of random variables, one wishes to assess the relationship between two variables, conditioned on the remaining variables. In graphical models, graphs are used to display the conditional independence structure of the variables.

Consider a graph ${\cal G} = \left( V, {\cal E} \right)$ with a set of $p$ vertices (nodes) $V = \{1,2, \cdots , p\} =[p]$, and a corresponding set of (undirected) edges ${\cal E} \subset [p] \times [p]$. Given a (real-valued) random vector ${\bm x} = [x_1 \; x_2 \; \cdots \; x_p]^\top$, in the corresponding graph ${\cal G}$, each variable $x_i$ is represented by a node ($i$ in $V$), and associations between variables $x_i$ and $x_j$ are represented by edges between nodes $i$ and $j$ of ${\cal G}$. In a conditional independence graph (CIG), there is no edge between nodes $i$ and $j$ if and only if (iff) $x_i$ and $x_j$ are conditionally independent given the remaining $p-2$ variables. Gaussian graphical models (GGM) are CIGs where ${\bm x}$ is real-valued multivariate Gaussian. Suppose ${\bm x}$ has positive-definite covariance matrix $\bm{\Sigma}$ ($= \mathbb{E} \{ {\bm x} {\bm x}^\top \}$) with inverse covariance matrix (also known as precision matrix or concentration matrix) $\bm{\Omega} = \bm{\Sigma}^{-1}$. Then ${\Omega}_{ij}$, the $(i,j)-$th element of  $\bm{\Omega}$, is zero iff $x_i$ and $x_j$ are conditionally independent \cite{Whittaker1990, Lauritzen1996, Buhlmann2011}. 

Graphical models were originally developed for random vectors with multiple independent realizations, i.e., for time series that is independent and identically distributed (i.i.d.). Such models have been extensively studied \cite{Danaher2014, Friedman2004, Lauritzen2003, Meinshausen2006, Mohan2014}. Graphical modeling of real-valued time-dependent data (stationary time series) originated with \cite{Brillinger1996}, followed by \cite{Dahlhaus2000}. Time series graphical models of i.i.d.\ or dependent data have been applied to intensive care monitoring \cite{Gather2002}, financial time series \cite{Abdelwahab2008, Songsiri2009, Songsiri2010, Khare2015}, social networks \cite{Goldenberg2009}, air pollution data \cite{Dahlhaus2000, Songsiri2009}, analysis of EEG \cite{Medkour2010, Wolstenholme2015, Luftman2016}, and fMRI (functional magnetic resonance imaging) data \cite{Marrelec2006, Songsiri2010, Ryali2012}, colon tumor classification \cite{Rothman2008} and breast cancer data analysis \cite{Lam2009}. A significant technical issue in these analyses and applications is that of model selection. Given $p$ nodes in $V$, in an undirected graph, there are $p(p-1)/2$ distinct edges. Which edges are in ${\cal E}$, and which are not -- this is the model selection (graph learning) problem. 

Now consider a stationary (real-valued), zero-mean,  $p-$dimensional multivariate Gaussian time series ${\bm x}(t)$, $t=0, \pm 1, \pm 2, \cdots $, with $i$th component $x_i(t)$. In a dependent time series GGM ${\cal G}$, edge $\{i,j\} \in {\cal E}$ iff time series components $\{ x_i(t) \}$ and $\{ x_j(t) \}$ are conditionally dependent. In \cite{Dahlhaus2000} the term ``partial correlation graph'' is used for such graphs. A key insight in \cite{Dahlhaus2000} was to transform the series to the frequency domain and express the graph relationships in the frequency domain. Denote the power spectral density (PSD) matrix of $\{ {\bm x}(t) \}$ by ${\bm S}_x(f)$, where ${\bm S}_x(f) = \sum_{\tau = -\infty}^{\infty}  {\bm R}_{xx}( \tau ) e^{-j 2 \pi f \tau}$, the Fourier transform of ${\bm R}_{xx}( \tau ) = \mathbb{E} \{ {\bm x}(t + \tau) {\bm x}^T(t ) \}$.  Here $f$ is the normalized frequency, in Hz. Since for real-valued time series, ${\bm S}_x(f) = {\bm S}_x^H(-f)$, and ${\bm S}_x(f)$ is periodic in $f$ with period one, knowledge of ${\bm S}_x(f)$ in the interval $[0,0.5]$ completely specifies ${\bm S}_x(f)$ for other values of $f$. In \cite{Dahlhaus2000} it was shown that conditional independence of two time series components given all other components of the time series, is encoded by zeros in the inverse PSD, that is, $\{ i,j \} \not\in {\cal E}$ iff the $(i,j)$-th element of ${\bm S}_x^{-1}(f)$,  $[{\bm S}_x^{-1}(f)]_{ij} = 0$ for every $f$. This paper is concerned with sparse high-dimensional graphical modeling of dependent time series. It is noted in  \cite{Dahlhaus2000} that for partial correlation graph estimation via nonparametric methods, checking for $[{\bm S}_x^{-1}(f)]_{ij} \equiv 0$ is computationally much less demanding than using time-domain methods where one would  need to calculate $2 {p \choose 2}$ linear filters (see \cite[p.\ 161]{Dahlhaus2000} for details). 

Comparing the facts that $\{ i,j \} \not\in {\cal E} \; \Leftrightarrow \; {\Omega}_{ij} =0$ for vector GGMs, while $\{ i,j \} \not\in {\cal E} \; \Leftrightarrow \; [{\bm S}_x^{-1}(f)]_{ij} = 0 \; \forall \, f \in [0,0.5]$ for dependent time series GGMs, we see that ${\bm S}_x^{-1}(f)$ plays the same role for a dependent time series as is done by the concentration matrix $\bm{\Omega}$ in the i.i.d.\ time series (i.e., random vector) setting. The normalized DFT ${\bm d}_x(f_m)$, $f_m = m/n$, of real-valued time-domain data $\{ {\bm x}(t) \}_{t=0}^{n-1}$, defined in (\ref{app1eq50}), plays a central role in our proposed approach, and use of DFT can also be viewed as a way to decorrelate time-domain dependent data by transforming it to frequency-domain where $\{{\bm d}_x(f_m)\}_{m=0}^{n/2}$ is an approximately independent sequence (see Sec.\ \ref{BPF1}). Representing  data $\{ {\bm x}(t) \}_{t=0}^{n-1}$ as well as the DFT sequence $\{{\bm d}_x(f_m)\}_{m=0}^{n-1}$ as column vectors, the two can be related via a unitary $n \times n$ matrix, signifying that $\{{\bm d}_x(f_m)\}_{m=0}^{n-1}$ represents any $\{ {\bm x}(t) \}_{t=0}^{n-1}$ via an orthonormal basis in $\mathbb{C}^n$ \cite[p.\ 71]{Stoica1997}. One may also view this transformation (linear projection) as feature extraction from raw data, more suitable than raw data for further processing for some intended task. In our case this transformation is invertible. In other applications, not necessarily related to graphical modeling, one may resort to combined projection and dimension reduction to a lower rank subspace; possible examples include \cite{Chang2016, Yuan2021} and references therein. Low rank bilinear projections of matrices (or higher-order tensors) are proposed in \cite{Chang2016} to improve the performance of existing two-dimensional discriminant analysis algorithms for classification tasks. A self-supervised learning method is proposed in \cite{Yuan2021} to train a deep feature extraction network without the need for numerous manually labeled video samples. It is not yet clear how such ideas and approaches apply to graphical model learning for dependent time series.

\subsection{Related Work}
Prior work on graphical modeling for dependent time series in the low-dimensional settings (sample size $n \gg p$) is concerned with testing whether $\{ i,j \} \in {\cal E}$ for all possible edges in the graph, based on some nonparametric frequency-domain test statistic such as partial coherence \cite{Dahlhaus2000, Gather2002, Matsuda2006, Medkour2010, Wolstenholme2015, Luftman2016, Tugnait19d} which, in turn, is based on estimates of ${\bm S}_x(f)$ given time-domain data. These approaches do not scale to high dimensions where $p$ is comparable to or larger than the sample size $n$. As an alternative to nonparametric modeling of time series, parametric graphical models utilizing a (Gaussian) vector autoregressive (VAR) process model of ${\bm x}(t)$, have been advocated in \cite{Eichler2006, Eichler2012, Songsiri2009}, but these approaches are suitable for only small values of $p$. As an alternative to exhaustive search over various edges, a penalized maximum likelihood approach in conjunction with VAR models has been used in \cite{Songsiri2010} where the penalty term incorporates sparsity constraints, making it suitable for high-dimensional setting. For every pair of series components, the corresponding partial coherence is thresholded to decide if it is zero (exclude the edge), or nonzero (include the edge). No systematic (or principled) method is given for  threshold selection.  

Nonparametric approaches for graphical modeling of real time series in high-dimensional settings have been formulated in frequency-domain in \cite{Jung2014, Jung2015b} using a neighborhood regression scheme, and in the form of penalized log-likelihood in frequency-domain in \cite{Jung2015a, Tank2015, Foti2016}, all based on estimates of ${\bm S}_x(f)$ at various frequencies, given time-domain data. A key model assumption used in these papers is that ${\bm S}_x^{-1}(f), \; f \in [0,1],$ is sparse, i.e., for any $f$, the number of off-diagonal nonzero elements are much smaller than the total number $p(p-1)$ of off-diagonal elements. Sparsity is enforced during optimization of the chosen objective function, making the problem well-conditioned. Reference \cite{Foti2016} considers latent variable graphical models. References \cite{Tank2015, Foti2016} exploit the framework of \cite{Jung2015a}, and therefore, inherit some of its drawbacks, discussed in more detail later in Sec.\ \ref{compare} (see also Remark 1 in Sec.\ \ref{BPF3}). Reference  \cite{Tank2015} considers a Bayesian framework with a focus on multiple time series realizations whereas \cite{Jung2015a} deals with a single realization of the time series. Sufficient conditions for graph edge recovery are provided in \cite{Jung2015a} whereas there is no such analysis in \cite{Tank2015, Foti2016}. A recent approach \cite{Tran2020} considers a vector Gaussian time series with uncorrelated samples but possibly nonstationary covariance matrices. This paper \cite{Tran2020} addresses real-valued time series using a neighborhood regression scheme. However, the results in \cite{Tran2020} are applied to the discrete Fourier transform (DFT) of stationary time series without analyzing the resultant complex-values series (complex in the frequency-domain), and without noting the fact the DFT over the entire observation set does not result in (approximately) uncorrelated Gaussian sequence; see Secs.\ \ref{BPF1} and \ref{compare} later where it is pointed out that only ``half'' of the DFT sequence is uncorrelated. 

\subsection{Our Contributions} \label{contrib}
In this paper we address the same problem as in \cite{Jung2015a}, namely, first estimate the inverse PSD ${\bm S}_x^{-1}(f)$ at distinct frequencies, given time-domain data, and then select the graph edge $\{i,j\}$ based on whether or not $[{\bm S}_x^{-1}(f)]_{ij} = 0$ for every $f$. As in \cite{Jung2015a}, we use a penalized log-likelihood function in frequency-domain as our objective function to estimate ${\bm S}_x^{-1}(f)$, given time-domain data. However, there are significant differences between our choice of frequencies and objective function, and our analysis, and that of \cite{Jung2015a}. We enumerate these differences below and elaborate on them some more later in the paper after introducing some concepts and notation in Sec.\ \ref{BPF}.

\begin{itemize}
\item[(i)] Our log-likelihood function is different from that in \cite{Jung2015a}. In terms of the DFT ${\bm d}_x(f_m)$, $f_m = m/n$, of real-valued time-domain data $\{ {\bm x}(t) \}_{t=0}^{n-1}$, defined in (\ref{app1eq50}), we establish that $\{ {\bm d}_x(f_m) \}_{m=0}^{n/2}$, ($n$ even), is a sufficient statistic for our problem, a fact not recognized in \cite{Jung2015a} (see also Remark 1 in Sec.\ \ref{BPF3}, and Sec.\ \ref{compare}), who uses some redundant frequencies $f_m$'s from the set $(\frac{1}{2},1)$, in addition to those from the set $[0,\frac{1}{2}]$. Our proposed frequency-domain formulation is based on $\{ {\bm d}_x(f_m) \}_{m=1}^{(n/2)-1}$, neglecting ${\bm d}_x(f_m)$ at $m=0$ and $m=n/2$ where the DFT is real-valued Gaussian. We use a sparse-group lasso penalty on off-diagonal elements of inverse PSD estimates at individual frequencies (lasso) as well as jointly across all frequencies (group lasso), whereas \cite{Jung2015a} uses only a group lasso penalty that is applied to all elements, including diagonal elements, of inverse PSD estimates.  The objective of lasso/group lasso penalty is to drive possible zero entries of inverse PSD matrix to zero. However, since PSD matrix at any frequency is assumed to be positive-definite, its inverse is also positive-definite, hence, all diagonal elements of inverse PSD matrix at any frequency are strictly positive, therefore, do not need any sparsity penalty. Our sparse-group lasso penalty is more general than the group lasso penalty of \cite{Jung2015a} and it reduces to group lasso penalty if we remove the lasso penalty at individual frequencies. In our optimization approach, we allow for the possibility of zero lasso penalty.

\item[(ii)] In \cite{Jung2015a}, a test statistic based on the estimated inverse PSD's is compared to a nonzero threshold to infer presence/absence of edges in the graph where no method is given regarding how to choose the threshold. In this paper the threshold is set to zero. We also empirically investigate selection of the tuning parameters (lasso and group-lasso penalty parameters) needed for our sparse-group lasso approach, based on the Bayesian information criterion (BIC). No such results are available in \cite{Jung2015a}.

\item[(iii)] We provide sufficient conditions (Theorem 1 in Sec.\ \ref{consist}) for convergence in the Frobenius norm of the inverse PSD estimators to the true value, jointly across all frequencies, where the number of frequencies are allowed to increase with sample size so that the entire sufficient statistic set $\{ {\bm d}_x(f_m) \}_{m=1}^{(n/2)-1}$ is exploited. This results also yields a rate
of convergence of the inverse PSD estimators with sample size $n$. Reference \cite{Jung2015a} provides sufficient conditions for consistent graph edge recovery in \cite[Proposition III.2]{Jung2015a} when the number of frequencies are fixed independent of sample size,  and \cite{Tank2015, Foti2016} offer no theoretical analysis. A consequence of using fixed number of frequencies is that in \cite{Jung2015a}, only a subset of sufficient statistic set is exploited; we elaborate on this later in Remark 1 in Sec.\ \ref{BPF3}. We do not have any counterpart to \cite[Proposition III.2]{Jung2015a} whereas \cite{Jung2015a} does not have any counterpart to our Theorem 1.

\item[(iv)] We propose an alternating direction method of multipliers (ADMM) approach for optimization of the sparse-group lasso-based penalized log-likelihood formulation of the problem. Note that \cite{Jung2015a} also used ADMM but only for group lasso formulation. We discuss practical ADMM issues such as choice of ADMM penalty parameter and stopping rule for termination of the algorithm whereas there is no such discussion in \cite{Jung2015a}.

\end{itemize} 

\subsection{Relationship to Prior Conference Publications} Preliminary versions of parts of this paper appear in conference papers \cite{Tugnait18c, Tugnait20}. The sufficient statistic for our problem was first discussed in \cite{Tugnait18c} where a sparse-group lasso problem was also formulated in frequency-domain. But an alternating minimization (AM) based solution to this problem, using the penalty method, is given in \cite{Tugnait18c}. The performance of this method depends upon the penalty parameter, and strictly speaking, convergence of this solution to the desired solution requires the penalty parameter to become large, which can make the problem numerically ill-conditioned. Use of the ADMM approach mitigates the dependence on the penalty parameter used in the AM approach. The theoretical analysis (Theorem 1) presented in Sec.\ \ref{consist} is partially given in \cite{Tugnait20} where the proof is incomplete. In this paper we provide a complete proof and also correct some typos. In turn, the proof of Theorem 1 relies on some prior results in \cite{Tugnait19c} which deal with complex-valued Gaussian vectors, not time series. Lemma 1 appears in \cite{Tugnait18c} without a proof. Lemma 3 is from \cite{Tugnait20} where a complete proof is given, here we simply state and use it. A version of Lemma 4 is stated without proof in \cite{Tugnait19c}, here we give a complete proof. Lots of details of proof of Theorem 1 are missing in \cite{Tugnait20}.  The material in Secs.\ \ref{stopping}, \ref{BIC} and \ref{examples} of this paper does not appear in \cite{Tugnait18c, Tugnait19c, Tugnait20}.

\subsection{Outline and Notation}  
The rest of the paper is organized as follows. The sufficient statistic in the frequency-domain for our problem and the resulting log-likelihood formulation of the problem are presented in Sec.\  \ref{BPF} where we also provide some background material on Wirtinger calculus needed for optimization w.r.t.\ complex variables, and Whittle likelihood used in \cite{Jung2015a}. A sparse-group lasso-based penalized log-likelihood formulation of the problem is introduced in  Sec.\  \ref{PLL}.
An ADMM algorithm is presented in Sec.\  \ref{opt} to optimize the objective function to estimate the inverse PSD and the edges
in the graph. Selection of the tuning parameters based on BIC is presented in Sec.\ \ref{BIC}. In Sec.\ \ref{consist} we analyze consistency (Theorem 1)  of the proposed approach. Numerical results based on synthetic as well as real data are presented in Sec.\ \ref{examples} to illustrate the proposed approach. We present numerical results for both synthetic and real data. In synthetic data example the ground truth is known and this allows for assessment of the efficacy of various approaches. In real data example where the ground truth is unknown, our goal is visualization and exploration of the dependency structure underlying the data. In both cases we wish to compare our proposed approach with the widely used i.i.d.\ modeling approach where the underlying time series is either assumed to be i.i.d., or one uses only the covariance of the data. Proofs of Lemma 1 and Theorem 1 are given in Appendices \ref{proofL1} and \ref{proof}, respectively.

The superscripts $\ast$, $\top$ and $H$ denote the complex conjugate, transpose and  Hermitian (conjugate transpose) operations, respectively, and the sets of real and complex numbers are denoted by $\mathbb{R}$ and $\mathbb{C}$, respectively.
Given  ${\bm A} \in \mathbb{C}^{p \times p}$, we use $\phi_{\min }({\bm A})$, $\phi_{\max }({\bm A})$, $|{\bf A}|$, $\mbox{tr}({\bm A})$ and $\mbox{etr}({\bm A})$ to denote the minimum eigenvalue, maximum eigenvalue, determinant, trace, and exponential of trace of ${\bm A}$, respectively. We use ${\bm A} \succeq 0$ and ${\bm A} \succ 0$ to denote that Hermitian ${\bm A}$ is positive semi-definite and positive definite, respectively. The Kronecker product of matrices ${\bm A}$ and ${\bm B}$ is denotes by ${\bm A} \otimes {\bm B}$. For ${\bm B} \in \mathbb{C}^{p \times q}$, we define  the operator norm, the Frobenius norm and the vectorized $\ell_1$ norm, respectively, as $\|{\bm B}\| = \sqrt{\phi_{\max }({\bm B}^H  {\bm B})}$, $\|{\bm B}\|_F = \sqrt{\mbox{tr}({\bm B}^H  {\bm B})}$ and $\|{\bm B}\|_1 = \sum_{i,j} |B_{ij}|$, where $B_{ij}$ is the $(i,j)$-th element of ${\bm B}$, also denoted by $[{\bm B}]_{ij}$. 
For vector ${\bm \theta} \in \mathbb{C}^p$, we define $\| {\bm \theta} \|_1 = \sum_{i=1}^p |\theta_i|$ and $\| {\bm \theta} \|_2 = \sqrt{\sum_{i=1}^p |\theta_i|^2}$, and we also use $\| {\bm \theta} \|$ for $\| {\bm \theta} \|_2$. Given ${\bm A} \in \mathbb{C}^{p \times p}$, ${\bm A}^+ = \mbox{diag}({\bm A})$ is a diagonal matrix with the same diagonal as ${\bm A}$, and  ${\bm A}^- = {\bm A} - {\bm A}^+$ is ${\bm A}$ with all its diagonal elements set to zero. We use ${\bm A}^{-\ast}$ for $({\bm A}^\ast)^{-1}$, the inverse of complex conjugate of ${\bm A}$, and ${\bm A}^{- \top}$ for $({\bm A}^\top)^{-1}$. Given ${\bm A} \in \mathbb{C}^{n \times p}$, column vector $\mbox{vec}({\bm A}) \in \mathbb{C}^{np}$ denotes the vectorization of ${\bm A}$ which stacks the columns of the matrix ${\bm A}$. The notation ${\bm y}_n = {\cal O}_P({\bm x}_n)$ for random ${\bm y}_n, {\bm x}_n \in \mathbb{C}^p$ means that for any $\varepsilon > 0$, there exists $0 < M < \infty$ such that $P ( \|{\bm y}_n\| \le M \|{\bm x}_n\|) \ge 1 - \varepsilon$ $\forall n \ge 1$. The notation ${\bm x} \sim {\mathcal N}_c( {\bf m}, {\bm \Sigma})$ denotes a complex random vector  ${\bm x}$ that is circularly symmetric (proper), complex Gaussian with mean ${\bm m}$ and covariance ${\bm \Sigma}$, and ${\bm x} \sim {\mathcal N}_r( {\bf m}, {\bm \Sigma})$ denotes real-valued Gaussian ${\bm x}$ with mean ${\bm m}$ and covariance ${\bm \Sigma}$. The abbreviations PSD, w.r.t., w.h.p., h.o.t., iff and pdf stand for power spectral density, with respect to, with high probability, higher-order terms, if and only if, and probability density function, respectively.

\section{Background and Problem Formulation} \label{BPF}
Given time-domain data $\{ {\bm x}(t) \}_{t=0}^{n-1}$ originating from a $p-$dimensional stationary Gaussian sequence, our objective is to first estimate the inverse PSD ${\bm S}_x^{-1}(f)$ at distinct frequencies, and then select the edge $\{i,j\}$ in the time series GGM ${\cal G}$ based on whether or not $[{\bm S}_x^{-1}(f)]_{ij} = 0$ for every $f$. We will follow a maximum likelihood approach, and to this end we need to express the likelihood function of time-domain data in terms its PSD matrix.

\subsection{Sufficient Statistic and Log-Likelihood} \label{BPF1}
Given ${\bm x}(t)$ for $t=0, 1,2, \cdots , n-1$. Define the (normalized) DFT ${\bm d}_x(f_m)$ of ${\bm x}(t)$, ($j = \sqrt{-1}$),
\begin{equation} \label{app1eq50}
   {\bm d}_x(f_m) = \frac{1}{\sqrt{n}} \sum_{t=0}^{n-1} {\bm x}(t) \exp \left( - j 2 \pi f_m t \right) 
\end{equation}
where
\begin{equation} \label{app1eq50a}
	   f_m = m/n, \quad m=0,1, \cdots , n-1.
\end{equation}
Since $\{ {\bm x}(t) \}$ is Gaussian, so is ${\bm d}_x(f_m)$. Note that ${\bm d}_x(f_m)$ is periodic in $m$ with period $n$, and is periodic in normalized frequency $f_m$ with period 1. Since ${\bm x}(t)$ is real-valued, we have ${\bm d}_x^\ast(f_m) = {\bm d}_x(-f_m) = {\bm d}_x(1-f_m)$, so ${\bm d}_x(f_m)$ for $m=0,1, \cdots , (n/2)$, ($n$ even), completely determines ${\bm d}_x(f_m)$ for all integers $m$.
As proved in \cite[p.\ 280, Sec.\ 6.2]{Casella}, for any statistical inference problem, the complete sample is a sufficient statistic, and so is any one-to-one function of a sufficient statistic.  Since the inverse DFT yields (one-to-one transformation)
${\bm x}(t) =  \frac{1}{\sqrt{n}}\sum_{m=0}^{n-1} {\bm d}_x(f_m) e^{j 2 \pi f_m t}$,
the set $\{{\bm d}_x(f_m)\}_{m=0}^{n-1}$ is a sufficient statistic, which can be further reduced to $\{{\bm d}_x(f_m)\}_{m=0}^{n/2}$ since ${\bm x}(t)$ is real-valued, inducing symmetries ${\bm d}_x^\ast(f_m) = {\bm d}_x(-f_m) = {\bm d}_x(1-f_m)$. Thus, the set of complex-valued random vectors $\{{\bm d}_x(f_m)\}_{m=0}^{n/2}$ is a sufficient statistic for any statistical inference problem, including our problem of estimation of inverse PSD.

We need the following assumption in order to invoke \cite[Theorem 4.4.1]{Brillinger}, used extensively later.
\begin{itemize}
\setlength{\itemindent}{0.1in}
\item[(A1)] The $p-$dimensional time series $\{ {\bm x}(t) \}_{t=-\infty}^{\infty}$ is zero-mean stationary and Gaussian, satisfying 
\[
    \sum_{\tau = -\infty}^\infty | [{\bm R}_{xx}( \tau )]_{k \ell} | < \infty \mbox{ for every } 
		  k, \ell \in V  \, .
\] 
\end{itemize}
It follows from \cite[Theorem 4.4.1]{Brillinger} that under assumption (A1), asymptotically (as $n \rightarrow \infty$), ${\bm d}_x(f_m)$, $m=1,2, \cdots , (n/2)-1$, ($n$ even), are independent proper (i.e., circularly symmetric), complex Gaussian ${\mathcal N}_c( {\bf 0}, {\bm S}_x(f_m))$ random vectors, respectively. Also, asymptotically, ${\bm d}_x(f_0)$ and ${\bm d}_x(f_{n/2})$, ($n$ even), are independent real Gaussian ${\mathcal N}_r( {\bf 0}, {\bm S}_x(f_0))$ and ${\mathcal N}_r( {\bf 0}, {\bm S}_x(f_{n/2}))$ random vectors, respectively, independent of ${\bm d}_x(f_m)$, $m \in \{ 1, 2, \cdots, (n/2)-1 \}$. We will ignore these two frequency points $f_0$ and $f_{n/2}$.

Define 
\begin{equation} \label{eqth1_150a}
    {\bm D} = \left[ {\bm d}_x(f_1) \;  \cdots \; {\bm d}_x(f_{(n/2)-1}) \right]
	   \in \mathbb{C}^{p \times ((n/2)-1)} \, .
\end{equation}
Under assumption (A1), the asymptotic joint probability density function (pdf) of ${\textbf D}$ is given by
\begin{equation} \label{eqth1_150}
	 f_{{\bm D}}({\textbf D})  =  \prod_{m=1}^{(n/2)-1} 
	\frac{ \exp \left(-{\bm d}_x^H(f_m) {\bm S}_x^{-1}(f_m) 
								 {\bm d}_x(f_m) \right) }{\pi^p \, |{\bm S}_x(f_m)|} \, ,
\end{equation}
leading to the log-likelihood function
\begin{align} \label{eqth1_150b}
	 \ln f_{{\bm D}}({\bm D})  = & - \sum_{m=1}^{(n/2)-1} \big(
	   \ln |{\bm S}_x(f_m)|  \nonumber \\
			& + {\bm d}_x^H(f_m) {\bm S}_x^{-1}(f_m) 
								 {\bm d}_x(f_m) \big) - p(\frac{n}{2}-1) \ln \pi  \, .
\end{align}

\subsection{Complex Gaussian Vectors}  \label{BPF1a}
Here we recall some facts regarding proper and improper complex Gaussian random vectors from \cite{Schreier10}. We need these results to clarify different expressions for the pdf of a proper complex Gaussian vector, used later for optimization w.r.t.\ complex variables using Wirtinger calculus \cite[Appendix 2]{Schreier10}, \cite{Hjorungnes07}. Define ${\bm R}_{vw} = \mathbb{E}\{ {\bm v} {\bm w}^\top\}$ for (zero-mean) ${\bm v}, {\bm w} \in \mathbb{R}^p$, and define the covariance matrix ${\bm R}_{vw} = \mathbb{E}\{ {\bm v} {\bm w}^H\}$, and the complementary covariance matrix $\tilde{\bm R}_{vw} = \mathbb{E}\{ {\bm v} {\bm w}^\top\}$ \cite[Sec.\ 2.2]{Schreier10}, for zero-mean ${\bm v}, {\bm w} \in \mathbb{C}^p$.
Given ${\bm u} = {\bm u}_r + j {\bm u}_i \in \mathbb{C}^p$, with real part ${\bm u}_r$ and imaginary part ${\bm u}_i$, define the augmented complex vector ${\bm y}$ and the real vector ${\bm z}$ as
\begin{equation}  \label{aeq005}
    {\bm y} = \left[ \begin{array}{cc} {\bm u}^\top & {\bm u}^H \end{array} \right]^\top , \quad
    {\bm z} = \left[ \begin{array}{cc} {\bm u}_r^\top & {\bm u}_i^\top \end{array} \right]^\top  .
\end{equation}
The pdf of an improper complex Gaussian ${\bm u}$ is defined in terms of that of the augmented  ${\bm z}$ or ${\bm y}$ \cite[Sec.\ 2.3.1]{Schreier10}. Assume $\mathbb{E}\{ {\bm u} \} = {\bm 0}$. Then we have ${\bm z} \sim {\cal N}_r({\bm 0}, {\bm R}_{zz})$ where
\begin{equation}  \label{aeq010}
    {\bm R}_{zz} = \left[ \begin{array}{cc} {\bm R}_{u_r u_r} & {\bm R}_{u_r u_i} \\ 
		{\bm R}_{u_i u_r} & {\bm R}_{u_i u_i} \end{array} \right]  \, , \;
		{\bm R}_{yy}  = \left[ \begin{array}{cc} 
		 {\bm R}_{uu} & \tilde{\bm R}_{uu} \\
          \tilde{\bm R}_{uu}^\ast & {\bm R}_{uu}^\ast  \end{array} \right] 
					  = {\bm R}_{yy}^H \, .
\end{equation}

Since ${\bm z} \sim {\cal N}_r({\textbf 0}, {\bm R}_{zz})$, its pdf is given by (assuming ${\bm R}_{zz} \succ {\textbf 0}$)
\begin{align}
   f_{{\bm z}}({\bm z}) =&  \frac{1}{ (2 \pi)^{2p/2} \, |{\bm R}_{zz}|^{1/2}} 
			          \exp \left(- \frac{1}{2} {\bm z}^\top {\bm R}_{zz}^{-1} {\bm z} \right) \, .  \label{aeq030}
\end{align}
One can also express (\ref{aeq030}) as \cite[Sec.\ 2.3.1]{Schreier10}
\begin{align}
   f_{{\bm u}}({\bm u}) & :=  f_{{\bm y}}({\bm y}) 
	  = \frac{1}{ \pi^p \, |{\bm R}_{yy}|^{1/2}} 
			\exp \left(- \frac{1}{2} {\bm y}^H {\bm R}_{yy}^{-1} {\bm y} \right) \, .  \label{aeq035}
\end{align}
For proper complex ${\bm u}$, $\tilde{\bm R}_{uu} = {\bm 0}$, and (\ref{aeq035}) reduces to 
\begin{align}
   f_{{\bm u}}({\bm u}) & = \frac{ e^{- \frac{1}{2} {\bm u}^H {\bm R}_{uu}^{-1} {\bm u} 
			     - \frac{1}{2} \left( {\bm u}^H {\bm R}_{uu}^{-1} {\bm u} \right)^\ast } }
					 { \pi^p \, |{\bm R}_{uu}|^{1/2} \, |{\bm R}_{uu}^\ast|^{1/2} }  \, .  \label{aeq036}
\end{align}
Since ${\bm R}_{uu} = {\bm R}_{uu}^H$, $|{\bm R}_{uu}| = |{\bm R}_{uu}^\ast|$, it follows that $( {\bm u}^H {\bm R}_{uu}^{-1} {\bm u} )^\ast = {\bm u}^H {\bm R}_{uu}^{-1} {\bm u}$, for proper ${\bm u}$, and therefore, (\ref{aeq036}) has the familiar form used in (\ref{eqth1_150}).

\subsection{Wirtinger Calculus}  \label{BPF1b}
In this paper we will optimize a scalar objective function of complex-values matrices. So we will use Wirtinger calculus (complex differential calculus) \cite[Appendix 2]{Schreier10}, \cite{Hjorungnes07}, coupled with corresponding definition of subdifferential/subgradients \cite{Li2015, Ollila2016}, to analyze and minimize a strictly convex objective function of complex-values matrices, e.g., function $L_{SGL} (\{ \bm{\Phi} \})$ of complex $\{ \bm{\Phi} \}$ defined in (\ref{eqth2_20}). We will use the necessary and sufficient Karush-Kuhn-Tucker (KKT) conditions for a global optimum. Consider a complex-valued ${\bm z} = {\bm x} +j {\bm y} \in \mathbb{C}^p$, ${\bm x},{\bm y}$ reals, and a real-valued scalar function $g({\bm z}) = g({\bm z},{\bm z}^\ast) = g({\bm x},{\bm y})$. In Wirtinger calculus, one views $g({\bm z})$ as a function $g({\bm z},{\bm z}^\ast)$ of two independent vectors ${\bm z}$ and ${\bm z}^\ast$, instead of a function a single ${\bm z}$, and defines
\begin{align}
 \frac{\partial g({\bm z},{\bm z}^\ast)}{\partial {\bm z}^\ast} := &\frac{1}{2} \left[ \frac{\partial g}{\partial {\bm x}} + j \frac{\partial g}{\partial {\bm y}} \right] \\
  \frac{\partial g({\bm z},{\bm z}^\ast)}{\partial {\bm z}} := & \frac{1}{2} \left[ \frac{\partial g}{\partial {\bm x}} - j \frac{\partial g}{\partial {\bm y}} \right] \, ;
\end{align}
see \cite[Appendix 2]{Schreier10}. For $g({\bm z})$ one defines its subdifferential $\partial g({\bm z}_0)$ at a point  ${\bm z}_0$ as \cite{Li2015, Ollila2016}
\begin{align}
   \partial g({\bm z}_0) = & \Big\{ {\bm s} \in \mathbb{C}^p \, : \,  g({\bm z}) \ge g({\bm z}_0)
	     + 2 \, \mbox{Re}\left( {\bm s}^H ({\bm z}-{\bm z}_0)  \right)  \nonumber \\
		& \quad\quad \mbox{ for all } {\bm z} \in \mathbb{C}^p \Big\} \, .
\end{align}
In particular, for scalar $z \in \mathbb{C}$, $g(z) =|z|$, we have \cite{Ollila2016}
\begin{align}
  2\, \partial |z| & = t = \left\{ \begin{array}{ll}
			    z/|z| & \mbox{if } z \neq 0 \\
					\in \{ v \,:\, |v| \le 1, \; v \in \mathbb{C} \} & \mbox{if } z = 0 \end{array} \right. .
\end{align}
Similarly, with $h_k(x) := g(z_1,z_2, \cdots , z_{k-1}, x, z_{k+1}, \cdots , z_p)$, $x \in \mathbb{C}$, the partial subdifferential $\partial g_{z_{0k}}({\bm z}) := \partial h_k(z_{0k})$ is the subdifferential $\partial h_k(z_{0k})$ of $h_k(x)$ at $z_{0k}$. 
Also \cite{Ollila2016}
\begin{equation}
   \partial g({\bm z}_0) = \frac{\partial g({\bm z})}{\partial {\bm z}^\ast} \Big|_{{\bm z}={\bm z}_0}
\end{equation}
when this partial derivative exists and $g$ is convex.

\subsection{Whittle Likelihood} \label{BPF2} Define ${\bm y} = [{\bm x}^\top(0) \; {\bm x}^\top(1) \; \cdots \; {\bm x}^\top(n-1)]^\top \in \mathbb{R}^{pn}$. By assumption ${\bm y} \sim {\cal N}_r({\bm 0}, {\bm \Sigma}_y)$ where ${\bm \Sigma}_y = E \{ {\bm y} {\bm y}^\top \}$. Since ${\bm \Sigma}_y \succ {\bm 0}$, the pdf of ${\bm y}$ is given by 
\begin{equation} \label{cpp1eq50}
   f_{{\bm y}}({\bm y}) = \frac{1}{ (2 \pi )^{np/2} |{\bm \Sigma}_y|^{1/2}} \exp (-{\bm y}^\top {\bm \Sigma}_y^{-1} {\bm y}) \,. 
\end{equation}
Based on some large sample ($n \rightarrow \infty$) results of Whittle \cite{Whittle1953, Whittle1953b, Whittle1957},  {\it Whittle approximation} to $f_{{\bm y}}({\bm y})$ is stated in \cite[Eqn.\ (5)]{Bach2004} and \cite[Eqn.\ (1)]{Rosen2007} as follows. \cite[Eqn.\ (1)]{Rosen2007} is
\begin{equation} \label{cpp1eq52}
   f_{{\bm y}}({\bm y}) \approx \prod_{m=0}^{n-1} 
	\frac{ \exp \left(-{\bm d}_x^H(f_m) {\bm S}_x^{-1}(f_m) 
								 {\bm d}_x(f_m) \right) }{ |{\bm S}_x(f_m)|} \, , 
\end{equation}
which specifies the joint pdf up to some constants, while \cite[Eqn.\ (5)]{Bach2004} specifies 
\begin{align} \label{cpp1eq54}
   \ln f_{{\bm y}}({\bm y}) \approx & - \frac{1}{2} \sum_{m=0}^{n-1} \big(
	   \ln |{\bm S}_x(f_m)|  \nonumber \\
			& + \mbox{tr}( {\bm S}_x^{-1}(f_m)  {\bm d}_x(f_m) {\bm d}_x^H(f_m)  ) \big) - \frac{pn}{2}  \ln (2 \pi )  \, , 
\end{align}
up to some constants. As noted in Sec.\ \ref{BPF1}, the terms in (\ref{cpp1eq52}) and (\ref{cpp1eq54}) corresponding to the indices $m=\frac{n}{2}+1$ through $n-1$ are completely specified by terms corresponding to the indices $m=0$ through $\frac{n}{2}$. For instance, ${\bm d}_x(f_{(n/2)+1}) =  {\bm d}_x^\ast(1-f_{(n/2)+1}) = {\bm d}_x^\ast(f_{(n/2)-1})$. Therefore, unlike (\ref{eqth1_150}) and (\ref{eqth1_150b}), (\ref{cpp1eq52}) and (\ref{cpp1eq54}), respectively, have lots of redundant frequencies. We note that \cite{Jung2015a} uses a likelihood function based on such Whittle approximation. The likelihood of \cite{Jung2015a} is examined further in Remark 1 in Sec.\ \ref{BPF3}.

\subsection{Problem Formulation} \label{BPF3}
Recall that our objective is to first estimate the inverse PSD ${\bm S}_x^{-1}(f)$ at distinct frequencies, and then select the edge $\{i,j\}$ in the time series GGM ${\cal G}$ based on whether or not $[{\bm S}_x^{-1}(f)]_{ij} = 0$ for every $f$. Suppose to obtain a maximum-likelihood estimate of inverse PSD ${\bm S}_x^{-1}(f)$, we minimize $-\ln f_{{\bm D}}({\bm D})$ with respect to ${\bm \Phi}_m := {\bm S}_x^{-1}(f_m)$. Then the problem is separable in $m$, and we choose ${\bm \Phi}_m$ to minimize 
\begin{align}
 h({\bm \Phi}_m) & := -\ln |{\bm \Phi}_m|  + {\bm d}_x^H(f_m) {\bm \Phi}_m {\bm d}_x(f_m) \nonumber \\
   & = -\frac{1}{2} \ln |{\bm \Phi}_m| - \frac{1}{2} \ln |{\bm \Phi}_m^\ast| \nonumber \\
	  & \;\; + \frac{1}{2} \mbox{tr} ({\bm \Phi}_m {\bm d}_x(f_m) {\bm d}_x^H(f_m)  + {\bm \Phi}_m^{\ast} {\bm d}_x^\ast(f_m) {\bm d}_x^\top(f_m))
\end{align}
where the expression after equality above follows by specifying the pdf of ${\bm d}_x(f_m)$ in terms of joint pdf of ${\bm d}_x(f_m)$ and ${\bm d}_x^\ast(f_m)$ as in (\ref{aeq036}) (correct way to handle complex variates \cite{Schreier10}).  Using Wirtinger calculus (Sec.\ \ref{BPF1b}), at the optimal solution the gradient of $h({\bm \Phi}_m)$ w.r.t.\ ${\bm \Phi}_m^\ast$ vanishes:
\begin{align}
 {\bm 0} & = \frac{\partial h({\bm \Phi}_m)}{\partial {\bm \Phi}_m^\ast} \nonumber \\
  & = - \frac{1}{2} ({\bm \Phi}_m^H)^{-1} + \frac{1}{2} ({\bm d}_x^\ast(f_m) {\bm d}_x^\top(f_m))^\top \nonumber \\
   & = - \frac{1}{2} {\bm \Phi}_m^{-1} + \frac{1}{2} {\bm d}_x(f_m) {\bm d}_x^H(f_m)
\end{align}
leading to the solution ($\hat{{\bm \Phi}}_m$ denotes estimate of ${{\bm \Phi}}_m$)
\begin{align}
   \hat{{\bm \Phi}}_m^{-1} = {\bm d}_x(f_m) {\bm d}_x^H(f_m) \, .
\end{align}
Since ${\bm d}_x(f_m) {\bm d}_x^H(f_m)$ is rank one, we cannot obtain $\hat{\bm \Phi}_m$ by inverting ${\bm d}_x(f_m) {\bm d}_x^H(f_m)$. Indeed, ${\bm d}_x(f_m) {\bm d}_x^H(f_m)$ is the periodogram \cite{Stoica1997} and it is known to be a poor estimator of the PSD ${\bm S}_x(f_m) ={\bm \Phi}_m^{-1}$. To obtain a consistent PSD estimator of ${\bm S}_x(f_m)$, one needs to smooth (average) ${\bm d}_x(f_k) {\bm d}_x^H(f_k)$ over values of $k$ centered around $m$, either directly (periodogram smoothing) or indirectly (Blackman-Tukey estimators operating on estimated correlation function) \cite[Chapter 2]{Stoica1997}, \cite[Sec.\ II-D.2]{Bach2004}. Under high dimensional case, one also needs sparsity constraints in order to regularize the problem.

We assume that ${\textbf S}_x(f_m)$ is locally smooth (a standard assumption in PSD estimation \cite[Chapter 2]{Stoica1997}, \cite{Brillinger}), so that ${\textbf S}_x(f_m)$ is (approximately) constant over $K=2m_t+1$ consecutive frequency points. Pick
\begin{equation} \label{window}
  \tilde{f}_k = \frac{(k-1)K+m_t+1}{n}, \;\; \quad k=1,2, \cdots , M \, ,
\end{equation}
\begin{equation} \label{windowM}
	 M = \left\lfloor \frac{ \frac{n}{2}-m_t-1}{K} \right\rfloor \, ,
\end{equation}
leading to $M$ equally spaced frequencies $\tilde{f}_k$ in the interval $(0,0.5)$, at intervals of $K/n$. It is assumed that for each $\tilde{f}_k$ (local smoothness),
\begin{equation}  \label{eqth1_160}
   {\textbf S}_x(\tilde{f}_{k,\ell}) = {\textbf S}_x(\tilde{f}_k) \,  
	\; \mbox{for  } \ell = -m_t, -m_t+1, \cdots , m_t,
\end{equation}
where
\begin{equation}  \label{eqth1_160a}
	  \tilde{f}_{k,\ell} = \frac{(k-1)K+m_t+1 + \ell}{n}.
\end{equation}

Using (\ref{eqth1_160}) in (\ref{eqth1_150}), we have 
\begin{align}  
   f_{{\textbf D}}({\textbf D})  = & \prod_{k=1}^{M}  \left[ \prod_{\ell = -m_t}^{m_t}
	   \frac{\exp \left(-{\textbf d}_x^H(\tilde{f}_{k,\ell}) {\textbf S}_x^{-1}(\tilde{f}_k) 
			      {\textbf d}_x(\tilde{f}_{k,\ell}) \right) }{ \pi^p \, | {\textbf S}_x(\tilde{f}_k) | }
		     \right] \, .  \label{eqth1_162} 
\end{align}
Define
\begin{align}  
   \check{\textbf D}(\tilde{f}_k) = & \left[ {\textbf d}_x(\tilde{f}_{k,-m_t}) \; {\textbf d}_x(\tilde{f}_{k,-m_t+1}) \; \cdots \; {\textbf d}_x(\tilde{f}_{k,m_t}) \right]^H \, , \\
	\tilde{\textbf D}(\tilde{f}_k) = & \sum_{\ell= - m_t}^{m_t} {\textbf d}_x(\tilde{f}_{k,\ell}) 
		 {\textbf d}_x^H(\tilde{f}_{k,\ell}) \, , \\ 
	\hat{\bm S}_k = &  \frac{1}{K} \tilde{\textbf D}(\tilde{f}_k)		\label{moresm}
\end{align}
where $\hat{\bm S}_k$ represents PSD estimator at frequency $\tilde{f}_{k}$ using unweighted frequency-domain smoothing \cite{Brillinger}.  The joint pdf of $\check{\textbf D}(\tilde{f}_k)$ is given by 
\begin{align}  
   f_{\check{\bm D}(\tilde{f}_k)}(\check{\bm D}(\tilde{f}_k))   = & \frac{ \exp \left(- {\rm tr} ( \tilde{\textbf D}(\tilde{f}_k) 
			       {\textbf S}_x^{-1}(\tilde{f}_k) ) \right) }{\pi^{Kp} \, | {\textbf S}_x(\tilde{f}_k) |^K }		\\ 
		= &  \frac{ \exp \left(- {\rm tr} ( K \hat{\bm S}_k 
			       {\textbf S}_x^{-1}(\tilde{f}_k) ) \right) }{\pi^{Kp} \, | {\textbf S}_x(\tilde{f}_k) |^K }	\, . \label{eqth1_163} 
\end{align}
Then we can rewrite (\ref{eqth1_162} ) as
\begin{align}
	f_{{\textbf D}}({\textbf D})  = & \prod_{k=1}^{M} f_{\check{\bm D}(\tilde{f}_k)}(\check{\bm D}(\tilde{f}_k))		\\ 
		= & \prod_{k=1}^{M} \frac{ \exp \left(- {\rm tr} ( K \hat{\bm S}_k 
			       {\textbf S}_x^{-1}(\tilde{f}_k) ) \right) }{\pi^{Kp} \, | {\textbf S}_x(\tilde{f}_k) |^K }	\, .
						\label{eqth1_164}
\end{align}
This is the joint pdf we will use in the rest of the paper.

{\it Remark 1}. Ref.\ \cite{Jung2015a} cites the Whittle approximation (\ref{cpp1eq54}) given in \cite{Bach2004} as a basis for their negative log likelihood function, which can be inferred from \cite[Eqn.\ (6)]{Jung2015a} after removing the lasso penalty therein.  In the notation of this paper, it is given by
\begin{align}
	- \ln f_{{\bm y}}({\bm y})  \propto 
		 & \sum_{k=1}^{F} \left( {\rm tr} (  \hat{\bm S}_k {\bm X}_k - \ln | {\bm X}_k | \right)	
						\label{eqj10}
\end{align}
where $\hat{\bm S}_k$ is the PSD matrix estimator at frequency $f_k = \frac{k-1}{F}$, $k=1,2, \cdots , F$, and ${\bm X}_k$ parametrizes the unknown inverse PSD ${\bm S}_x^{-1}(f_k)$. Minimization of penalized $- \ln f_{{\bm y}}({\bm y})$ w.r.t.\ ${\bm X}_k$, $k=1,2, \cdots , F$, yields the estimates of  ${\bm S}_x^{-1}(f_k)$, which are then used to infer the underlying graph. We will now relate (\ref{eqj10}) with $F$ frequencies, to (\ref{cpp1eq54}) with $n$ frequencies. In the analysis of \cite{Jung2015a}, $F$ is kept fixed while the sample size $n$ is allowed to increase. The estimate $\hat{\bm S}_k$ is obtained via Blackman-Tukey method which is mathematically equivalent to frequency-domain weighted smoothing of periodgram  ${\bm P}_m := {\bm d}_x(f_m) {\bm d}_x^H(f_m)$ for values of $m$ in a neighborhood of $k$, i.e., 
\begin{align}
	\hat{\bm S}_k \approx \sum_{\ell= - W}^{W} w_\ell {\bm d}_x(f_{k+\ell}) {\bm d}_x^H(f_{k+\ell})	
	            = \sum_{\ell= - W}^{W} w_\ell {\bm P}_{k+\ell} 
						\label{eqj15}
\end{align} 
where the weights $w_\ell$ and effective width $W$ depend upon the window function used in the time-domain on estimated correlation function (see \cite[Eqn.\ (8)]{Jung2015a} and also discussion in \cite[Sec.\ II-D.2]{Bach2004} and \cite[Sec.\ 2.5.1]{Stoica1997}).  A standard assumption is that of local smoothness of the true PSD matrix ${\bm S}_x(f_k)$, hence of ${\bm X_k}$, as in (\ref{eqth1_160}). This allows us to rewrite (\ref{eqj10}) as
\begin{align}
		 & \sum_{k=1}^{F} \left( {\rm tr} (  \hat{\bm S}_k {\bm X}_k - \ln | {\bm X}_k | \right)	\nonumber \\
		& \quad \approx
		  \sum_{k=1}^{F} \sum_{\ell= - W}^{W} \left( {\rm tr} ( w_\ell {\bm P}_{k+\ell} {\bm X}_{k+\ell} )
			             - \ln | {\bm X}_{k+\ell} | \right) \, .          
						\label{eqj20}
\end{align}
The total number of DFT frequencies used in (\ref{eqj20}), hence in (\ref{eqj10}), is $(2W+1)F$ whereas that in (\ref{cpp1eq54}) is $n$. For $\hat{\bm S}_k$ to be a consistent estimator of ${\bm S}_x(f_k)$, one must have $2W+1 \rightarrow \infty$ and $\frac{2W+1}{n} \rightarrow 0$ as $n \rightarrow \infty$ \cite[Secs.\ 5.6 and 7.4]{Brillinger}:  $\frac{2W+1}{n} \rightarrow 0$ makes $\hat{\bm S}_k$ asymptotically unbiased and  $2W+1 \rightarrow \infty$ makes estimator covariance matrix tend to zero. To have the same number of frequencies in  (\ref{eqj10}) (or (\ref{eqj20})) and (\ref{cpp1eq54}), one must have $(2W+1)F \approx n$, i.e., $F \approx n/(2W+1)$ ($\rightarrow \infty$ for consistency), which is not possible for the approach of \cite{Jung2015a} with fixed $F$. Note that the analysis of \cite{Jung2015a} requires $\hat{\bm S}_k$ to be a consistent estimator of ${\bm S}_x(f_k)$, which is possible by letting $2W+1 \rightarrow \infty$ and $\frac{2W+1}{n} \rightarrow 0$. But with fixed $F$, the product $(2W+1)F \ll n$ for large $n$ since $F$ is fixed. Thus, for large $n$, \cite{Jung2015a} exploits only a subset of the sufficient statistic set, whereas in our approach the product $ K M \approx \frac{n}{2}$ ($K$ and $M$ correspond to $2W+1$ and $F$, respectively, in (\ref{eqj20})), thereby using the entire sufficient statistic set. As noted in Sec.\ \ref{consist} (after (\ref{eqth2_20a})), in our case, as $n \rightarrow \infty$, we have $K \rightarrow \infty$, $M \rightarrow \infty$ and $\frac{K}{n} \rightarrow 0$.
$\quad \Box$

\section{Penalized Log-Likelihood} \label{PLL}
We wish to estimate inverse PSD matrix $\bm{\Phi}_k := {\bm S}_x^{-1}(\tilde{f}_k)$. In terms of $\bm{\Phi}_k$ we rewrite (\ref{eqth1_164}) as
\begin{align}
   & f_{\check{\bm D}(\tilde{f}_k)}(\check{\bm D}(\tilde{f}_k)) = 
	  \frac{   | \bm{\Phi}_k |^K e^{-{\rm tr} ( K \hat{\bm S}_k \bm{\Phi}_k ) }}{\pi^{Kp}  } \nonumber \\
			& \quad = \frac{   | \bm{\Phi}_k |^{K/2} | \bm{\Phi}_k^\ast |^{K/2} }{\pi^{Kp}  } 
			 e^{- {\rm tr} \left(  \frac{K}{2} \big( \hat{\bm S}_k \bm{\Phi}_k + \hat{\bm S}_k^\ast \bm{\Phi}_k^\ast \big) \right)} 
			 \label{eqth1_164a}
\end{align}
where the last expression in (\ref{eqth1_164a}) follows by specifying the pdf of $\check{\bm D}$ in terms of joint pdf of $\check{\bm D}$ and $\check{\bm D}^\ast$ as in (\ref{aeq036}). Then we have the log-likelihood (up to some constants) \cite{Tugnait18c}
\begin{align} 
   & \ln  f_{{\bm D}}({\bm D})  \propto - G(\{ \bm{\Phi} \}, \{ \bm{\Phi}^\ast \}) \\
					&	:= \sum_{k=1}^M \frac{1}{2} \left[( \ln |\bm{\Phi}_k| +\ln |\bm{\Phi}_k^\ast| )
		        - {\rm tr} \left( \hat{\bm S}_k \bm{\Phi}_k  +
						     \hat{\bm S}_k^\ast \bm{\Phi}_k^\ast \right)  \right] \, .  	 \label{eqth2_10}
\end{align}

In the high-dimensional case of $K < p^2$ (number of real-valued unknowns in ${\bm S}_x^{-1}(\tilde{f}_k) )$), one has to use penalty terms to enforce sparsity and to make the problem well-conditioned. Consider minimization of a convex objective function $g({\bm \theta})$ w.r.t.\ ${\bm \theta} \in \mathbb{R}^p$. If ${\bm \theta}$ is known to be sparse (only a few nonzero entries), one may choose ${\bm \theta}$ to minimize a lasso-penalized cost $g({\bm \theta}) + \lambda \|{\bm \theta}\|_1$ where $\lambda >0$ is the lasso penalty or tuning parameter used to control sparsity of the solution \cite{Tibshirani1996}. Suppose ${\bm \theta}$ has $M$ groups (subvectors) ${\bm \theta}^{(m)}$, $m=1,2, \cdots , M$, where rather than just sparsity in ${\bm \theta}$, one would like a solution which uses only a few of the groups ${\bm \theta}^{(m)}$ (group sparsity). To this end, \cite{Yuan2007} proposed a group lasso penalty where ${\bm \theta}$ is chosen to minimize a group lasso penalized cost $g({\bm \theta}) + \lambda \sum_{m=1}^M \|{\bm \theta}^{(m)}\|_2$ and where $\lambda >0$ is the group-lasso penalty parameter. As noted in \cite{Friedman2010a, Friedman2010b}, while the group-lasso gives a sparse set of groups, if it includes a group in the model, then all coefficients in the group will be nonzero. To enforce sparsity of groups and within each group, sparse-group lasso framework has been proposed in  \cite{Friedman2010a, Friedman2010b} where ${\bm \theta}$ is chosen to minimize a sparse-group lasso penalized cost $g({\bm \theta}) +  \alpha \lambda \|{\bm \theta}\|_1 + (1-\alpha) \lambda \sum_{m=1}^M \|{\bm \theta}^{(m)}\|_2$ where $\alpha \in [0,1]$ provides a convex combination of lasso and group-lasso penalties. Note that $\alpha =0$ gives the group-lasso fit while $\alpha =1$ yields the lasso fit, thus sparse-group lasso penalty is more general than either lasso or group-lasso penalties.  

Lasso penalty has been used in \cite{Friedman2008, Banerjee2008, Rothman2008, Ravikumar2011}, group lasso has been used in \cite{Kolar2014} and sparse-group lasso has been used in \cite{Danaher2014, Tugnait21}, all for graphical modeling of real-valued random vectors (i.e., i.i.d.\ time series) in various contexts. Group lasso has been used in \cite{Jung2015a} for graphical modeling of dependent time series. Results of \cite{Friedman2010a, Friedman2010b} for a regression problem and that of \cite{Tugnait21} for graphical modeling of random vectors in a multi-attribute context (a random vector is associated with each node of a graph instead of just a random variable), both show significant performance improvements over either just lasso or just group lasso penalties. For our problem in this paper we will use sparse-group lasso penalty.

Imposing a sparse-group sparsity constraint, we propose to minimize a penalized version of negative log-likelihood w.r.t.\ $\{ \bm{\Phi} \} = \{\bm{\Phi}_k, \; k=1,2, \cdots , M\}$, given by $L_{SGL} (\{ \bm{\Phi} \})$,
\begin{align}  
   L_{SGL} (\{ \bm{\Phi} \}) & =  G(\{ \bm{\Phi} \}, \{ \bm{\Phi}^\ast \}) + P(\{ \bm{\Phi} \}) , \label{eqth2_20} \\
	   P (\{ \bm{\Phi} \} ) &  =  \alpha \lambda \, \sum_{k=1}^M \; \sum_{i \ne j}^p 
		  \Big| [ {\bm{\Phi}}_k ]_{ij} \Big|  
		    + (1-\alpha) \lambda \, \sum_{ i \ne j}^p \; \| {\bm{\Phi}}^{(ij)} \|  \label{eqth2_20b} 
\end{align}
where
\begin{align}  
{\bm{\Phi}}^{(ij)} & := [ [{\bm{\Phi}}_1 ]_{ij} \; [{\bm{\Phi}}_2 ]_{ij} \; \cdots \; [{\bm{\Phi}}_M ]_{ij}]^\top
     \in \mathbb{C}^M  \, , \label{eqth2_20c}
\end{align}
$\lambda > 0$ and $\alpha \in [0,1]$. In (\ref{eqth2_20b}), an $\ell_1$ penalty term is applied to each off-diagonal element of ${\bm{\Phi}}_k$ via $\alpha \lambda \, \Big| [ {\bm{\Phi}}_k ]_{ij} \Big|$ (lasso), and to the off-block-diagonal group of $M$ terms via $(1-\alpha) \lambda \sqrt{\sum_{k=1}^{M} | [{\bm{\Phi}}_k ]_{ij} |^2 } $ (group lasso). 

To optimize $L_{SGL}(\{ \bm{\Phi} \})$, using variable splitting, one may reformulate as in \cite{Tugnait18c}: 
\begin{align}  \label{eqth2_21}
 \min_{ \{\bm{\Phi} \}, \{ {\bm W} \} } & \Big\{ G(\{ \bm{\Phi} \}, \{ \bm{\Phi}^\ast \})
						+ P(\{ {\bm W} \})  \Big\} \;\;
\end{align}
subject to ${\bm W}_k = \bm{\Phi}_k \succ {\bf 0}, \; k=1,2, \cdots , M$, where $\{ {\bm W} \} = \{{\bm W}_k, \; k=1,2, \cdots , M \}$. Using the penalty method, \cite{Tugnait18c} considers the {\em relaxed} problem ($\rho > 0$ is ``large'')
\begin{align}  \label{eqth2_22}
 \min_{\small \substack{ \{\bm{\Phi} \}, \\ \{ {\bm W} \}} } & \left\{ G(\{ \bm{\Phi} \}, \{ \bm{\Phi}^\ast \})
						+ P(\{ {\bm W} \})   
						  + \frac{\rho}{2} \sum_{k=1}^M \| \bm{\Phi}_k - {\bm W}_k \|^2_F \right\} 
\end{align} 
where it is solved via an alternating minimization (AM) based method \cite{Beck2015}.  The final result depends on $\rho$ and strictly speaking, one must have $\rho \rightarrow \infty$ which can make the problem numerically ill-conditioned. In the numerical example of \cite{Tugnait18c}, a fixed value $\rho =10$ was considered which does not necessarily achieve the constraint $\bm{\Phi}_k - {\bm W}_k= {\bm 0}$. On the other hand, in the ADMM approach which also uses a penalty parameter $\rho$, the final solution to the optimization problem does not depend upon the chosen $\rho > 0$, although the convergence speed depends on it \cite{Boyd2010}. In this paper we consider ADMM. Some authors \cite{Zhang2016, Zheng2018} have suggested that when both AM and ADMM approaches are applicable, the ADMM which requires dual variables, is inferior to the AM method which is a primal-only method, in terms of computational complexity and accuracy, but their claims do not account for the need to solve the AM problem multiple times, each time with increased value of $\rho$, and moreover, they do not consider graphical modeling problems.

\section{Optimization via ADMM}  \label{opt}

In ADMM, we consider the scaled augmented Lagrangian for this problem \cite{Li2015, Boyd2010}, given by 
\begin{align}
 L_\rho(\{ \bm{\Phi} \}, \{{\bm W} \}, \{{\bm U} \} ) = &  
   G(\{ \bm{\Phi} \}, \{ \bm{\Phi}^\ast \})
						+ {P}(\{ {\bm W} \})  \nonumber  \\
					&	+ \frac{\rho}{2} \sum_{k=1}^M \| \bm{\Phi}_k - {\bm W}_k + {\bm U}_k \|^2_F
\end{align}
where $\{ {\bm U} \} = \{{\bm U}_k, \; k=1,2, \cdots , M \}$ are dual variables, and $\rho >0$ is the ``penalty parameter'' \cite{Boyd2010}. 

\subsection{ADMM Algorithm}
Given the results $\{ \bm{\Phi}^{(m)} \}, \{{\bm W}^{(m)} \}, \{{\bm U}^{(m)} \}$ of the $m$th iteration, in the $(m+1)$st iteration, an ADMM algorithm executes the following three updates:
\begin{itemize}[wide, labelwidth=!, labelindent=0pt]
\setlength{\itemindent}{0.04in}
\item[(a)] $\{ \bm{\Phi}^{(m+1)} \} \leftarrow \arg \min _{\{ \bm{\Phi} \}} 
            L_\rho ( \{ \bm{\Phi} \}, \{{\bm W}^{(m)} \}, \{{\bm U}^{(m)} \} )$
\item[(b)] $\{ {\bf W}^{(m+1)} \} \leftarrow \arg \min _{\{ {\bm W} \}} 
            L_\rho(\{ \bm{\Phi}^{(m+1)} \}, \{{\bm W} \}, \{{\bm U}^{(m)} \} )$
\item[(c)] $\{ {\bm U}^{(m+1)} \} \leftarrow \{ {\bm U}^{(m)} \}  +
   \left( \{ \bm{\Phi}^{(m+1)} \}  - \{ {\bm W}^{(m+1)} \} \right)$
\end{itemize}

\subsubsection{Update (a)} \label{updatea}
Notice that $L_\rho ( \{ \bm{\Phi} \}, \{{\bm W}^{(m)} \}, \{{\bm U}^{(m)} \} )$ is separable in $k$ with $L_\rho ( \{ \bm{\Phi} \}, \{{\bm W}^{(m)} \}, \{{\bm U}^{(m)} \} ) = \sum_{k=1}^M \frac{1}{2} L_{\rho k} (  \bm{\Phi}_k , {\bm W}_k^{(m)}, {\bm U}_k^{(m)} )$ up to some terms not dependent upon $\bm{\Phi}_k$'s, where
\begin{align}  
	   L_{\rho k} & (  \bm{\Phi}_k , {\bm W}_k^{(m)}, {\bm U}_k^{(m)} )  :=  - \ln |\bm{\Phi}_k| -\ln |\bm{\Phi}_k^\ast| 
		        + {\rm tr} \Big( \hat{\bm S}_k \bm{\Phi}_k  \nonumber \\
					& \quad + 	     \hat{\bm S}_k^\ast \bm{\Phi}_k^\ast \Big)
	  + \rho \| \bm{\Phi}_k - {\bm W}_k^{(m)} + {\bm U}_k^{(m)} \|^2_F \, .
\end{align}
As in \cite[Sec.\ 6.5]{Boyd2010} but accounting for complex-valued vectors/matrices in this paper compared to real-valued vectors/matrices in \cite{Danaher2014}, and therefore using the Wirtinger calculus, the solution to $\arg \min_{ \bm{\Phi}_k} 
            L_{\rho k} (  \bm{\Phi}_k , {\bm W}_k^{(m)}, {\bm U}_k^{(m)} )$ is as follows. A necessary and sufficient condition for a global optimum is that the gradient of $L_{\rho k} (  \bm{\Phi}_k , {\bm W}_k^{(m)}, {\bm U}_k^{(m)} )$ w.r.t.\ $\bm{\Phi}_k^\ast$, given by (\ref{eqth2_2000}), vanishes, with $\bm{\Phi}_k = \bm{\Phi}_k^H \succ {\textbf 0}$ (we set ${\bm A} = {\bm W}_k^{(m)} - {\bm U}_k^{(m)}$) :
\begin{align}  
   {\bf 0} & = \frac{\partial L_{\rho k} (  \bm{\Phi}_k , {\bm W}_k^{(m)}, {\bm U}_k^{(m)} ) }{\partial \bm{\Phi}_k^\ast}  \nonumber \\
	   & = - (\bm{\Phi}_k^H)^{-1} + \hat{\bm S}_k^H + \rho ( \bm{\Phi}_k - {\bm A}) \label{eqth2_2000a} \\
		 & =  \hat{\bm S}_k - 
	       \bm{\Phi}_k^{-1} + \rho ( \bm{\Phi}_k - {\bm A}) \, . \label{eqth2_2000}
\end{align}
The solution to (\ref{eqth2_2000}) follows as for the real-valued case discussed in \cite[Sec.\ 6.5]{Boyd2010}. Rewrite (\ref{eqth2_2000}) as
\begin{equation}  \label{eqth2_2001}
     \hat{\bm S}_k - \rho  {\bm A}  = \hat{\bm S}_k - \rho ({\bm W}_k^{(m)} - {\bm U}_k^{(m)})  = \bm{\Phi}_k^{-1} - \rho \bm{\Phi}_k \, . 
\end{equation}						
Let ${\bm V}{\bm D}{\bm V}^H$ denote the eigen-decomposition of the Hermitian matrix $ \hat{\bm S}_k - \rho  {\bm A}$ where ${\bm D}$ is diagonal with real values on the diagonal, and ${\bm V}{\bm V}^H = {\bm V}^H{\bm V} = {\bm I}$. Then we have
\begin{equation}  \label{eqth2_2002}
     {\bm D}  = {\bm V}^H ( \bm{\Phi}_k^{-1} - \rho \bm{\Phi}_k ) {\bm V} 
		         = \tilde{\bm{D}}^{-1} - \rho \tilde{\bm{D}}
\end{equation}
where $\tilde{\bm{D}} := {\bm V}^H{\bm \Phi}_k{\bm V}$. Assume that $\tilde{\bm{D}}$ is a diagonal matrix and solve (\ref{eqth2_2002}) for diagonal $\tilde{\bm{D}}$. That is, $\tilde{ D}_{\ell \ell}$ should satisfy
\begin{equation}  \label{eqth2_2003}
     { D}_{\ell \ell} = 1/\tilde{D}_{\ell \ell} - \rho \tilde{{D}}_{\ell \ell}\, .
\end{equation}
The solution
\begin{equation}  \label{eqth2_2004}
     \tilde{ D}_{\ell \ell} = \frac{1}{2 \rho} \left( -{ D}_{\ell \ell} + \sqrt{ |{ D}_{\ell \ell}|^2 + 4 \rho } \, \right) 
\end{equation}
satisfies (\ref{eqth2_2003}) and yields $\tilde{ D}_{\ell \ell} > 0$ for any $\rho > 0$. Therefore, so constructed $\tilde{\bm{D}} \succ 0$, and hence, $\bm{\Phi}_k = \bm{\Phi}_k^{(m+1)} = {\bm V} \tilde{\bm D} {\bm V}^H \succ 0$ satisfies (\ref{eqth2_2000}) and (\ref{eqth2_2001}).

\subsubsection{Update (b)} \label{updateb}
Update $\{ {\bm W}_k^{(m+1)} \}_{k=1}^M$ as the minimizer w.r.t.\ $\{ {\bm W} \}_{k=1}^M$ of 
\begin{equation} \label{penaltyopt}
  \frac{\rho}{2} \sum_{k=1}^M \| {\bm W}_k - ( \bm{\Phi}_k^{(m+1)} + {\bm U}_k^{(m)} ) \|^2_F + {P}(\{ {\bm W} \}) \, .
\end{equation}
Here we use Lemma 1 first stated (but not proved) in \cite{Tugnait18c}, and is based on the real-valued results of \cite{Friedman2010a}. Lemma 1 is proved in Appendix \ref{proofL1}. \\
{\it Lemma 1}. Given ${\bm a} \in \mathbb{C}^q$, $\lambda_i >0$ ($i=1,2$), $h(\bm{\theta}) = (1/2) \|{\bm a}- \bm{\theta}\|^2 + \lambda_1 \sum_{i=1}^q |\theta_i| + \lambda_2 \|\bm{\theta}\|$ is minimized w.r.t.\ $\bm{\theta} \in \mathbb{C}^q$ by $\hat{\bm{\theta}}$ with the $i$th component
\begin{equation} \label{lemsol}
    \hat{\theta}_i = \left( 1- \frac{\lambda_2}{\|{\bm S}({\bm a}, \lambda_1)\|} \right)_+ S(a_i,\lambda_1)
\end{equation}
where $(b)_+ := \max(0,b)$, soft-thresholding operator $S(b,\beta) := (1-\beta/|b|)_+ b$ (for complex scalar $b \neq 0$), and vector operator $[{\bm S}({\bm a}, \beta)]_j = S(a_j,\beta)$, $a_j = [{\bm a}]_j$.
 $\;\; \bullet$ 

Define ${\bm A}_k = \bm{\Phi}_k^{(m+1)} + {\bm U}_k^{(m)}$ and let ${\bm A}^{(ij)} \in \mathbb{C}^M$ be defined as in (\ref{eqth2_20c}) but based on ${\bm A}_k$'s instead of ${\bm \Phi}_k$'s. Invoking Lemma 1, the solution to minimization of (\ref{penaltyopt}) is 
\begin{equation} \label{penaltysol1}
  [{\bm W}_k^{(m+1)}]_{ij} = \left\{ \begin{array}{l}
	    [{\bm A}_k]_{ii} \, , \quad\quad \mbox{if } i=j \\
			S([{\bm A}^{(ij)}]_{k}, \frac{\alpha \lambda}{\rho}) \left( 1 - \frac{(1-\alpha)\lambda}
			{ \rho \| {\bm S}({\bm A}^{(ij)}, \, \alpha \lambda/\rho)\|}
			   \right)_+ \\
				\quad\quad\quad\quad \mbox{if } i \neq j  \end{array} \right.
\end{equation}

\subsubsection{Update (c)} \label{updatec}
For the scaled Lagrangian formulation of ADMM \cite{Boyd2010}, for $k=1,2, \cdots , M$, update ${\bm U}_k^{(m+1)}={\bm U}_k^{(m)} +( \bm{\Phi}_k^{(m+1)} - {\bm W}_k^{(m+1)})$.

\subsubsection{Algorithm Outline} \label{outline}
\begin{itemize} %
\item[(i)] Initialize the variables: $\bm{\Phi}_k^{(0)} = {\textbf I}_p$, ${\bm W}_k^{(0)} = {\bm U}_k^{(0)} = {\bm 0}$ for $k=1,2, \cdots , M$. Pick scalar $\rho > 0$. 
\item[(ii)] Until convergence, for $m=1,2, \cdots , ...$, do steps (iii)-(v):
\begin{itemize}
\item[(iii)] For $k=1, \cdots , M$, update $\bm{\Phi}_k^{(m+1)}$ as in Sec.\ \ref{updatea}.
\item[(iv)] For $k=1, \cdots , M$, update ${\bm W}_k^{(m+1)}$ as in Sec.\ \ref{updateb}
\item[(v)] For $k=1, \cdots , M$, update ${\bm U}_k^{(m+1)}$ as in Sec.\ \ref{updatec}.
\end{itemize}
\item[(vi)] Denote the converged estimates as $\hat{\bm{\Phi}}_k$, $k=1, \cdots , M$. Edge selection:
\begin{equation} \label{thresh}
   \mbox{If } \| \hat{\bm \Phi}^{(ij)} \| \, > \, 0 , \mbox{ then } \{ i,j \} \in {\cal E} ,
	  \mbox{ else } \{ i,j \} \not\in {\cal E} .
\end{equation}
\end{itemize}

\subsubsection{Stopping Rule, Variable Penalty $\rho$, and Convergence}  \label{stopping}
{\it Stopping Rule}: In step (ii) of the Algorithm of Sec.\ \ref{outline}, we need a stopping (convergence) criterion to terminate the ADMM steps.  We will use a stopping criterion following \cite[Sec.\ 3.3.1]{Boyd2010}.  We also use a varying penalty parameter $\rho^{(m)}$ at the $m$th iteration, following \cite[Sec.\ 3.4.1]{Boyd2010}. The stopping criterion is based on primal and dual residuals of the ADMM approach being small. Minimization in (\ref{eqth2_21}) is done under the equality constraints ${\bm \Phi}_k - {\bm W}_k = {\bm 0}$, $k=1,2, \cdots , M$. The error in this equality during ADMM iterations is called primal residual as it measures the primal feasibility \cite[Sec.\ 3.3]{Boyd2010}. The primal residual matrix is given by 
\[
  {\bm R}_p = \begin{bmatrix} {\bm \Phi}_1 - {\bm W}_1, & \cdots , &  {\bm \Phi}_M - {\bm W}_M
           \end{bmatrix} \in \mathbb{C}^{p \times (pM)}
\]
and the primal residual vector is ${\bm r}_p = \mbox{vec} ({\bm R}_p) \in \mathbb{C}^{p^2 M}$, vectorization of ${\bm R}_p$. 
At the $(m+1)$st iteration, the primal residual matrix will be denoted by
\[
 {\bm R}_p^{(m+1)} = \begin{bmatrix} {\bm \Phi}_1^{(m+1)} - {\bm W}_1^{(m+1)}, & \cdots , &  {\bm \Phi}_M^{(m+1)} - {\bm W}_M^{(m+1)}
           \end{bmatrix}
\]
with corresponding vector ${\bm r}_p^{(m+1)} = \mbox{vec} ({\bm R}_p^{(m+1)})$. As $m \rightarrow \infty$, one must have ${\bm r}_p^{(m+1)} \rightarrow  0$. Based on some dual feasibility conditions for optimization of the ADMM problem, a dual residual at the $(m+1)$st iteration is defined in \cite[Sec.\ 3.3]{Boyd2010}. For our problem, the dual residual matrix at the $(m+1)$st iteration is given 
\[
  {\bm R}_d^{(m+1)} =  \rho^{(m)} \begin{bmatrix}  {\bm W}_1^{(m+1)} - {\bm W}_1^{(m)}, & \cdots , &
          {\bm W}_M^{(m+1)} - {\bm W}_M^{(m)} \end{bmatrix}  
\]
where ${\bm R}_d^{(m+1)} \in \mathbb{C}^{p \times (pM)}$, and the dual residual vector is ${\bm r}_d^{(m+1)} = \mbox{vec} ({\bm R}_d^{(m+1)}) \in \mathbb{C}^{p^2 M}$. As $m \rightarrow \infty$, one must have ${\bm r}_d^{(m+1)} \rightarrow  0$.

The convergence criterion is met when the norms of these residuals are below primary and dual tolerances $\tau_{pri}$ and $\tau_{dual}$, respectively: 
\begin{align*} d_p := & \|{\bm R}_p^{(m+1)}\|_F \le \tau_{pri} \\
d_d := & \|{\bm R}_d^{(m+1)}\|_F \le \tau_{dual} \, .
\end{align*} 
Following \cite[Sec.\ 3.3.1]{Boyd2010}, the tolerances $\tau_{pri}$ and $\tau_{dual}$ are chosen using an absolute and relative criterion based on user chosen absolute and relative tolerances $\tau_{abs}$ and $\tau_{rel}$. The absolute tolerance component of both $\tau_{pri}$ and $\tau_{dual}$ is $p \sqrt{M} \, \tau_{abs}$ where $p \sqrt{M}$ equals square-root of length of ${\bm r}_p^{(m+1)}$ as well as of ${\bm r}_d^{(m+1)}$. The relative tolerance components of $\tau_{pri}$ and $\tau_{dual}$ are proportional to the magnitude of the primary variables ${\bm \Phi}_k^{(m+1)}$ and ${\bm W}_k^{(m+1)}$, and dual variable ${\bm U}_k^{(m+1)}$, $k=1,2, \cdots , M$, respectively. Let
\begin{align*}
  e_1 = &\|[{\bm \Phi}_1^{(m+1)}, \cdots , {\bm \Phi}_M^{(m+1)}]\|_F \, , \\
	e_2 = &\|[{\bm W}_1^{(m+1)}, \cdots , {\bm W}_M^{(m+1)}]\|_F \, , \\
	e_3 = &\|[{\bm U}_1^{(m+1)}, \cdots , {\bm U}_M^{(m+1)}]\|_F \, .
\end{align*}
Then following \cite[Sec.\ 3.3.1]{Boyd2010}, we pick
\begin{align*}
   \tau_{pri} = & p \sqrt{M} \, \tau_{abs} + \tau_{rel} \, \max ( e_1, e_2 ) \\
  \tau_{dual} = & p \sqrt{M}  \, \tau_{abs} + \tau_{rel} \,  e_3 / \rho^{(m)} \, .
\end{align*}

{\it Variable Penalty $\rho$}: As stated in \cite[Sec.\ 3.4.1]{Boyd2010}, one may use ``possibly different penalty parameters $\rho^{(m)}$ for each iteration, with the goal of improving the convergence in practice, as well as making performance less dependent on the initial choice of the penalty parameter.''
For scaled Lagrangian formulation, the variable penalty $\rho$ is updated as \cite[Sec.\ 3.4.1]{Boyd2010}
\[
  \rho^{(m+1)} = \left\{ \begin{array}{ll} 2 \rho^{(m)} & \mbox{if  } d_p > \mu d_d \\  
	                                         \rho^{(m)} /2 & \mbox{if  } d_d > \mu d_p \\
																					 \rho^{(m)} & \mbox{otherwise}  \end{array} \right.
\]
for some $\mu > 1$.  As stated in \cite[Sec.\ 3.4.1]{Boyd2010}, ``The idea behind this penalty parameter update is to try to keep the primal and dual residual norms within a factor of $\mu$ of one another as they both converge to zero.'' This also necessitates a modification in step (v) of the Algorithm of Sec.\ \ref{outline}: After updating ${\bm U}_k^{(m+1)}$, further modify it as ${\bm U}_k^{(m+1)} ={\bm U}_k^{(m+1)}/2$ if $d_p > \mu d_d$, and ${\bm U}_k^{(m+1)} =2\, {\bm U}_k^{(m+1)}$ if $d_d > \mu d_p$ \cite[Sec.\ 3.4.1]{Boyd2010}. For all numerical results presented later, we used $\rho^{(0)} =2$ (initial value of $\rho$), $\mu =10$, and $\tau_{abs}=\tau_{rel} =10^{-4}$.

{\it Convergence}: The objective function $L_{SGL} (\{ \bm{\Phi} \})$, given by (\ref{eqth2_20}), is strictly convex in $\{ \bm{\Phi} \}$ for  ${\bm \Phi}_k  \succ {\bm 0}$, $k=1,2, \cdots , M$. It is also closed, proper and lower semi-continuous. Hence, for any fixed $\rho > 0$, the ADMM algorithm is guaranteed to converge \cite[Sec.\ 3.2]{Boyd2010}, in the sense that we have primal residual convergence to 0, dual residual convergence to 0, and objective function convergence to the optimal value. For varying $\rho$, the convergence of ADMM has not yet been proven \cite[Sec.\ 3.4.1]{Boyd2010}.

\subsection{Further on Comparison with Existing Works} \label{compare}
Here we briefly summarize comparisons with \cite{Jung2015a, Tank2015, Foti2016}. As noted earlier in Sec.\ \ref{BPF1}, $\{ {\bm d}_x(f_m) \}_{m=0}^{n/2}$ is a  frequency-domian sufficient statistic for this problem. However, in \cite{Jung2015a} (and \cite{Tank2015, Foti2016}), one uses some estimate of PSD ${\bm S}_x(f_m)$, or ${\bm d}_x(f_m)$, for $m=0,1, \cdots , n-1$ appealing to Whittle approximation which, as discussed in Sec.\ \ref{BPF2}, has lots of redundant frequencies. Also, in \cite{Jung2015a}, frequencies ($f_m$ or $\tilde{f}_k$ in our notation) are fixed a priori as $(f-1)/F \in [0,1)$, $f=1,2, \cdots , F$, for some even integer $F$. For instance, in the simulation example of \cite{Jung2015a}, $F=4$, leading to four frequencies $\{0, 0.25, 0.5, 0.75 \}$ for any data size $n$. Note that ${\bm S}_x(0.75) = {\bm S}_x^\ast(0.25)$ (and so are their estimates), so there is no new information in it. Also, ${\bm S}_x(0)$ and ${\bm S}_x(0.5)$, and their estimates, are real-valued, not complex, but as their estimates are based on ${\bm d}_x(f_k)$'s for $k$ in a neighborhood of $m=0$ or $0.5$, any information in the imaginary part of ${\bm d}_x(f_k)$'s is not exploited. Furthermore, as noted in item (i) in Sec.\ \ref{contrib}, \cite{Jung2015a} considers only group-lasso penalty which is subsumed by our more general sparse-group lasso. In our analysis presented later in Sec.\ \ref{consist}, the number of frequencies are allowed to increase with sample size, and as discussed in Remark 1, Sec.\ \ref{BPF3}, increasing the number of frequencies allows one to exploit the entire sufficient statistic set, unlike the analysis in \cite{Jung2015a} where with fixed $F$, one uses only a subset of sufficient statistics.

\subsection{BIC for Tuning Parameter Selection} \label{BIC}
Let $\hat{\bm{\Phi}}_k$, $k=1, \cdots , M$, denote the converged estimates, as noted in item (vi) of Sec.\ \ref{outline}.  
Given $n$ and choice of $K$ and $M$, the Bayesian information criterion (BIC) is given by
\begin{align}
  {\rm BIC}&(\lambda , \alpha) =  2K  \sum_{k=1}^M \left( -\ln |\hat{\bm{\Phi}}_k|  + {\rm tr} \left( \hat{\bm S}_k  \hat{\bm{\Phi}}_k  \right) \right)  \nonumber \\
	&  + \ln (2 K M) \, \sum_{k=1}^M (\mbox{\# of nonzero elements in } \hat{\bm{\Phi}}_k ) 	  
\end{align}
where $2KM$ are total number of real-valued measurements in frequency-domain and $2K$ are number of real-valued measurements per frequency point, with total $M$ frequencies in $(0, \pi)$. Each nonzero off-diagonal element of $\hat{\bm{\Phi}}_k$ consists of two real variables, but since $\hat{\bm{\Phi}}_k$ is Hermitian, the number of (nonzero) real unknowns in $\hat{\bm{\Phi}}_k$ equal the number of nonzero elements of $\hat{\bm{\Phi}}_k$. Pick $\alpha$ and $\lambda$ to minimize BIC. We use BIC to first select $\lambda$ over a grid of values with fixed $\alpha$, and then with selected $\lambda$, we search over $\alpha$ values in $[0,0.3]$. This sequential search is computationally less demanding than a two-dimensional search. 

We search over $\lambda$ in the range $[\lambda_\ell , \lambda_u]$ selected via the following heuristic (as in \cite{Tugnait21}). The heuristic lies in the fact that we limit the range of $\alpha$ and $\lambda$ values over which the search is performed. For $\alpha = \alpha_0$ (=0.1), we first find the smallest $\lambda$, labeled $\lambda_{sm}$, for which we get a no-edge model (i.e., $| \hat{\cal E} | =0$, where $\hat{\cal E}$ denotes the estimated edge set based on (\ref{thresh})). Then we set $\lambda_u = \lambda_{sm}/2$ and $\lambda_\ell = \lambda_u/10$. The given choice of $\lambda_u$ precludes ``extremely'' sparse models while that of $\lambda_\ell$ precludes ``very'' dense models. The choice $\alpha \in [0,0.3]$ reflects the fact that group-lasso penalty across all frequencies is more important ($(i,j)$th element of inverse PSD at all frequencies must be zero for edge $\{i,j\} \not\in {\cal E}$) than lasso penalty at individual frequencies. 

\section{Consistency} \label{consist}
In this section we analyze consistency of the proposed approach. In Theorem 1 we provide sufficient conditions for convergence in the Frobenius norm of the inverse PSD estimators to the true value, jointly across all frequencies, where the number of frequencies are allowed to increase with sample size. As discussed in Remark 1, Sec.\ \ref{BPF3}, increasing the number of frequencies allows one to exploit the entire sufficient statistic set, unlike the analysis in \cite{Jung2015a}. Theorem 1 also yields a rate of convergence w.r.t.\ sample size $n$. We follow proof technique of \cite{Rothman2008} which deals with i.i.d.\ time series models and lasso penalty, to establish our main result, Theorem 1.

Define $p \times (pM)$ matrix ${\bm \Omega}$ as
\begin{equation}  \label{neqn100}
  {\bm \Omega} =[ {\bm \Phi}_1 \;  {\bm \Phi}_2 \; \cdots \; {\bm \Phi}_M ] \, .
\end{equation}
With $0 \le \alpha \le 1$, re-express the objective function (\ref{eqth2_20}) as
\begin{align}  
   L_{SGL}(\bm{\Omega}) & =  G(\{ \bm{\Phi} \}, \{ \bm{\Phi}^\ast \}) 
	    + \alpha  \lambda_n \, \sum_{k=1}^{M_n} \; \sum_{\substack{i,j=1 \\ i \ne j}}^{p_n} 
		  \Big| [ {\bm{\Phi}}_k ]_{ij} \Big| \nonumber \\
		  &  \quad
		    + (1-\alpha ) \lambda_n \, \sum_{\substack{i,j=1 \\ i \ne j}}^{p_n} \; \sqrt{\sum_{k=1}^{M_n} 
				       \Big| [{\bm{\Phi}}_k ]_{ij} \Big|^2 }  \label{eqth2_20a}
\end{align}
where we now allow $p$, $M$, $K$ (see (\ref{window}), (\ref{windowM})), and $\lambda$ to be functions of sample size $n$, denoted as $p_n$, $M_n$, $K_n$ and $\lambda_n$, respectively. We take $p_n$ to be a non-decreasing function of $n$, as is typical in high-dimensional settings. Note that $K_n M_n \approx n/2$. Pick $K_n = a_1 n^\gamma$ and $M_n = a_2 n^{1-\gamma}$ for some $0.5 < \gamma < 1$, $0 < a_1, a_2 < \infty$, so that both $M_n/K_n \rightarrow 0$ and $K_n/n \rightarrow 0$ as $n \rightarrow \infty$ (cf.\ Remark 1). As discussed in Remark 2 later, Theorem 1 clarifies how to choose $M_n$ and $K_n$ (or put additional restrictions on it) so that for given $\{ p_n \}$, the estimate of ${\bm{\Omega}}$ converges to its true value in the Frobenius norm. 

Assume
\begin{itemize}
\setlength{\itemindent}{0.1in}
\item[(A2)] Define the true edge set of the graph by ${\cal E}_0$, implying that ${\cal E}_0 = \{ \{i,j\} ~:~ [{\bm S}^{-1}_{0}(f)]_{ij} \not\equiv 0, ~i\ne j,  ~ 0 \le f \le 0.5 \}$ where ${\bm S}_0(f) $ denotes the true PSD of ${\bm x}(t)$.
(We also use $\bm{\Phi}_{0k} $ for ${\bm S}^{-1}_{0}(\tilde{f}_k)$ where $\tilde{f}_k$ is as in (\ref{window}), and use ${\bm \Omega}_0$ to denote the true value of ${\bm \Omega}$). Assume that card$({\cal E}_0) =|({\cal E}_0)| \le s_{n0}$.

\item[(A3)] The minimum and maximum eigenvalues of $p_n \times p_n$ PSD ${\bm S}_0(f)  \succ {\bm 0}$  satisfy 
\begin{align*}
     0 < \beta_{\min} & \le \min_{f \in [0,0.5]} \phi_{\min}({\bm S}_0(f)) \\
		    & \le 
		     \max_{f \in [0,0.5]} \phi_{\max}({\bm S}_0(f)) \le \beta_{\max} < \infty \, .
\end{align*}
Here $\beta_{\min}$ and $\beta_{\max}$ are not functions of $n$ (or $p_n$).
\end{itemize}

Under assumptions (A1)-(A3), our main theoretical result is Theorem 1 stated below. Assumption (A1), stated in Sec.\ \ref{BPF1}, ensures that ${\bm S}_x(f)$ exists (\cite[Theorem 2.5.1]{Brillinger}) and it allows us to invoke \cite[Theorem 4.4.1]{Brillinger} regarding statistical properties of ${\bm d}_x(f_m)$ discussed in Sec.\ \ref{BPF1}. Assumption (A2) is more of a definition specifying the number of connected edges in the true graph to be upperbounded by $s_{n0}$. The maximum possible value of $s_{n0}$ is $p_n (p_n-1)$ (where we count edges $\{i,j\}$ and $\{j,i\}$, $i\ne j$, as two distinct edges), but we are interested in sparse graphs with $s_{n0} \ll p_n(p_n-1)$. The lower bound in assumption (A3) ensures that ${\bm S}_0^{-1}(f)$ exists for every $f \in [0,1]$. Existence of  ${\bm S}_0^{-1}(f)$ is central to this paper since its estimates are used to infer the underlying graph and it implies certain Markov properties of the conditional independence graph \cite[Lemma 3.1 and Theorem 3.3]{Dahlhaus2000}. The upper bound in assumption (A3) ensures that all elements in ${\bm S}_0(f)$ are uniformly upperbounded in magnitude. To prove this, first note that for any Hermitian ${\bm A} \in \mathbb{C}^{p \times p}$,  by \cite[Theorem 4.2.2]{Horn},
\begin{equation}  \label{naeq58e}
  \phi_{max}({\bm A}) = \max_{{\bm y} \ne {\bm 0}} \frac{ {\bm y}^H {\bm A} {\bm y} }{ {\bm y}^H  {\bm y} } \, .
\end{equation}
Consider ${\bm e}_{(i)} \in \mathbb{C}^p$ with one in the $i$th position and zeros everywhere else. Then  ${\bm e}_{(i)}^H {\bm A} {\bm e}_{(i)} = [{\bm A}]_{ii}$ and  ${\bm e}_{(i)}^H {\bm e}_{(i)} = 1$, leading to
\begin{equation}  \label{naeq58g}
  \phi_{max}({\bm A}) \ge  \frac{ {\bm e}_{(i)}^H {\bm A} {\bm e}_{(i)} }{ {\bm e}_{(i)}^H {\bm e}_{(i)} } = [{\bm A}]_{ii} \, .
\end{equation} 
Therefore, by (\ref{naeq58g}), for $1 \le i \le p_n$,
\begin{equation}  \label{naeq58h}
    [{\bm S}_0(f)]_{ii} \le \phi_{\max}({\bm S}_0(f)) \le \beta_{\max} \; \mbox{ for }  f \in [0,0.5] \, .
\end{equation}
Finally, by \cite[p.\ 279]{Brillinger}, 
\[
    |[{\bm S}_0(f)]_{k \ell}|^2 \le [{\bm S}_0(f)]_{kk} \, [{\bm S}_0(f)]_{\ell \ell}
\] 
implying that $|[{\bm S}_0(f)]_{k \ell}| \le \beta_{\max}$ for $f \in [0,0.5]$, $1 \le k, \ell \le p_n$.

Let $\hat{\bm{\Omega}}_\lambda = \arg\min_{\bm{\Omega} \,:\, \bm{\Phi}_{k} \succ {\bm 0}}  L_{SGL}(\bm{\Omega})$.
Theorem 1 whose proof is given in Appendix \ref{proof}, establishes consistency of $\hat{\bm{\Omega}}_\lambda$. \\
{\it Theorem 1 (Consistency)}. For $\tau > 2$, let
\begin{equation}  \label{naeq58}
   C_0 = 80 \, \max_{\ell, f} ( [{\bm S}_0(f)]_{\ell \ell}) 
	    \sqrt{ \frac{N_1}{ \ln ( p_n ) } } 
\end{equation}
where
\begin{equation}  \label{naeq58aa}
   N_1 = 2 \ln (16 p_n^\tau M_n ) \, .
\end{equation}
Given any real numbers $\delta_1 \in (0,1)$, $\delta_2 > 0$ and $C_1 > 0$, let
\begin{align}  
    R = & C_{2} C_0 / \beta_{\min}^2 , \quad 
		   C_{2} = 2(2+C_{1}+ \delta_2)(1+\delta_1)^2  \, , \label{neq15ab0} \\
    r_n = & \sqrt{ \frac{M_n( p_n+ s_{n0}) \ln (p_n )}{K_n}}, \quad C_0 C_2 r_n = o(1)\, , \label{neq15ab1} \\
		N_2 = & \arg \min \left\{ n \, : \, r_n \le 
		    \frac{ \delta_1 \beta_{\min}}{ C_2 C_0 } \right\} \, , \label{neq15ab3} \\
    N_3 = &  \arg \min \Big\{ n \, : \, K_n > N_1 \Big\} \, .  \label{neq15ab2} 
\end{align}
Suppose the regularization parameter $\lambda_n$ and $\alpha \in [0,1]$ satisfy 
\begin{align}  
    C_0 & \sqrt{\frac{\ln (p_n )}{K_n}}  \le \frac{\lambda_n}{\sqrt{M_n}}   \nonumber \\
		 & \le \frac{C_1  C_0 }{1+\alpha (\sqrt{M_n}-1)} 
		 \sqrt{ \left(1+ \frac{p_n}{s_{n0}} \right) \frac{\ln (p_n )}{K_n} } \, . \label{neq15abc}
\end{align}
Then if the sample size is such that $n >  \max \{ N_2, N_3 \}$ and assumptions (A1)-(A3) hold true,
$\hat{\bm{\Omega}}_\lambda$ satisfies
\begin{equation}  \label{neq15}
  \| \hat{\bm{\Omega}}_\lambda - \bm{\Omega}_0 \|_F 
	        \le R  r_n
\end{equation}
with probability greater than $1-1/p_n ^{\tau-2}$. In terms of rate of convergence (i.e., for large $n$),  
\begin{equation}  \label{neq1500}
	\| \hat{\bm{\Omega}}_\lambda - \bm{\Omega}_0 \|_F = {\cal O}_P \left( \ln (M_n) \sqrt{M_n} \, r_n  \right) \, .
\end{equation}
A sufficient condition for the lower bound in (\ref{neq15abc}) to be less than the upper bound for every $\alpha \in [0,1]$ is  $C_1 = 2(1+\alpha (\sqrt{M_n}-1))$. 
$\quad \bullet$
	
{\it Remark 2}. Theorem 1 helps determine how to choose $M_n$ and $K_n$ so that for given $\{ p_n \}$, $\lim_{n \rightarrow \infty} \| \hat{\bm{\Omega}}_\lambda - \bm{\Omega}_0 \|_F = 0$. This behavior is governed by  (\ref{neq1500}), therefore we have to examine $\ln (M_n) \sqrt{M_n} \, r_n$. As noted before, since $K_n M_n \approx n/2$, if one picks $K_n = a_1 n^\gamma$, then $M_n = a_2 n^{1-\gamma}$ for some $0 < \gamma < 1$, $0 < a_1, a_2 < \infty$. Suppose that the maximum number of nonzero elements in ${\bm S}_0^{-1}(f)$, $p_n+ s_{n0}$, satisfy $p_n+ s_{n0} = a_3 n^\theta$ for some $0 \le \theta < 1$, $0 < a_3 < \infty$. Then we have
\begin{align}
  {\cal O}  \left( \ln (M_n) \sqrt{M_n} r_n \right) = &
	   {\cal O} \left( \frac{\ln(M_n) M_n \sqrt{( p_n+ s_{n0}) \ln (p_n )}}{\sqrt{K_n}} \right) \nonumber \\
	        = & {\cal O} \left( \frac{(\ln(n))^{3/2} n^{1-\gamma} n^{\theta/2} }{n^{\gamma/2}} \right)  \nonumber \\
	        = & {\cal O} \left( \frac{(\ln(n))^{3/2}}{ n^{1.5 \gamma -1 - 0.5 \theta} } \right)  \label{neq1510} \\
					 \overset{n \uparrow \infty}{\rightarrow} & 0 
					 \mbox{ if } 1.5 \gamma -1 - 0.5 \theta >  0 \, .  \nonumber
\end{align}
If $\theta =0$ (fixed graph size and fixed number of connected edges w.r.t.\ sample size $n$), then we need $\frac{2}{3} < \gamma < 1$. By (\ref{neq1510}), we must have $1 > \gamma > \frac{2}{3} + \frac{\theta}{3}$. If $\theta > 0$, $\gamma$ has to be increased beyond what is needed for $\theta =0$, implying more smoothing of periodogram ${\bm d}_x(f_m) {\bm d}_x^H(f_m)$ around $f_k$ to estimate ${\bm S}_x(f_k)$ (recall (\ref{moresm})), leading to fewer frequency test points $M_n$. Clearly, we cannot have $\theta \ge 1$ because $p_n+ s_{n0} = {\cal O}(n^\theta)$ will require $\gamma > 1$ which is impossible.

If $\alpha =0$, then $C_1$ is a constant, and therefore, $\| \hat{\bm{\Omega}}_\lambda - \bm{\Omega}_0 \|_F = {\cal O}_P \left( \ln (M_n) r_n  \right)$. In this case we have 
\begin{align}
  {\cal O} \left( \ln (M_n) r_n \right) = & 
	   {\cal O} \left( \frac{\ln(M_n)  \sqrt{M_n ( p_n+ s_{n0}) \ln (p_n )}}{\sqrt{K_n}} \right) \nonumber \\
	        = & {\cal O} \left( \frac{(\ln(n))^{3/2} n^{(1-\gamma)/2} n^{\theta/2} }{n^{\gamma/2}} \right) \nonumber \\
	        = & {\cal O} \left(\frac{(\ln(n))^{3/2} }{ n^{(2 \gamma -1 -  \theta)/2} } \right) \label{neq1520} \\
					 \overset{n \uparrow \infty}{\rightarrow} & 0 
					 \mbox{ if } 2 \gamma -1 - \theta > 0 \, .  \nonumber
\end{align}
Now we must have $1 > \gamma > \frac{1}{2} + \frac{\theta}{2}$. If $\theta =0$, we need $\frac{1}{2} < \gamma < 1$. Also, we cannot have $\theta \ge 1$ because will require $\gamma > 1$. $\quad \Box$

\section{Numerical Examples}  \label{examples}
We now present numerical results for both synthetic and real data to illustrate the proposed approach. In synthetic data example the ground truth is known and this allows for assessment of the efficacy of various approaches. In real data example where the ground truth is unknown, our goal is visualization and exploration of the dependency structure underlying the data. 

\subsection{Synthetic Data} Consider $p=128$, 16 clusters (communities) of 8 nodes each, where nodes within a community are not connected to any nodes in other communities. Within any community of 8 nodes, the data are generated using a vector autoregressive (VAR) model of order 3. Consider community $q$, $q=1,2, \cdots , 16$. Then  ${\bm x}^{(q)}(t) \in \mathbb{R}^8$ is generated as
\[
    {\bm x}^{(q)}(t) = \sum_{i=1}^3 {\bm A}^{(q)}_i {\bm x}^{(q)}(t-i) + {\bm w}^{(q)}(t)
\] 
with  ${\bm w}^{(q)}(t)$ as i.i.d.\ zero-mean Gaussian with identity covariance matrix.
Only 10\% of entries of ${\bm A}^{(q)}_i$'s are nonzero and the nonzero elements are independently and uniformly distributed over $[-0.8,0.8]$. We then check if the VAR(3) model is stable with all eigenvalues of the companion matrix $\le 0.95$ in magnitude; if not, we re-draw randomly till this condition is fulfilled. The overall data ${\bm x}(t)$ is given by ${\bm x}(t) = [\, {\bm x}^{(1) \top}(t) \; \cdots \; {\bm x}^{(16) \top}(t) \, ]^\top \in \mathbb{R}^{p}$. First 100 samples are discarded to eliminate transients, and generate stationary Gaussian data. This set-up leads to approximately 3.5\% connected edges. In each run, we calculated the true PSD ${\bm S}(f)$ for $f \in [0,0.5]$ at intervals of 0.01, and then take $\{i,j\} \in {\cal E}$ if $\sum_f | S_{ij}^{-1}(f) | > 10^{-6}$. Note that average value of diagonal elements $(\sum_{i=1}^p \sum_f S_{ii}^{-1}(f))/p$, averaged over 100 runs, turns out to be 75.10 ($\pm 1.85$). Therefore, the threshold of $10^{-6}$ in $\sum_f | S_{ij}^{-1}(f) | > 10^{-6}$ is quite low, resulting in some very ``weak'' edges in the graph.

\begin{figure}[h]
  \centering
  \includegraphics[width=1.0\linewidth]{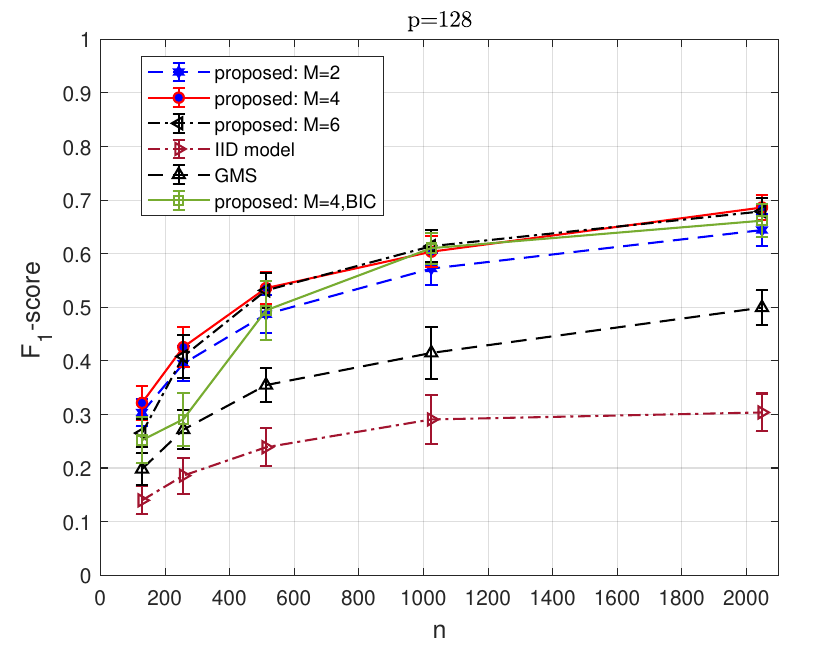} 
\caption{$F_1$-score for synthetic data example. The label ``GMS'' refers to the approach of \cite{Jung2015a}.}
\label{fig0}
\end{figure}

Simulation results are shown in Fig.\ \ref{fig0} where the performance measure is $F_1$-score for efficacy in edge detection. The $F_1$-score is defined as $F_1 = 2 \times \mbox{precision} \times \mbox{recall}/(\mbox{precision} + \mbox{recall})$ where $\mbox{precision} = | \hat{\cal E} \cap {\cal E}_0|/ |\hat{\cal E}|$, $\mbox{recall} = |\hat{\cal E} \cap {\cal E}_0|/ |{\cal E}_0|$, and ${\cal E}_0$ and $ \hat{\cal E}$ denote the true and estimated edge sets, respectively. For our proposed approach, we consider three different values of $M \in \{ 2,4 ,6 \}$ for five samples sizes $n \in \{128, 256, 512, 1024, 2048 \}$. For $M=2$, we used $K=31,63, 127, 255, 511 $ for $n=128,256,512,1024, 2048$, respectively, for $M=4$, we used $K=15,31,63,127, 255$ for $n=128,256,512,1024, 2048$, respectively, and for $M=6$, we used $K=9, 21, 41, 85, 169$ for $n=128,256,512,1024, 2048$, respectively. These approaches are labeled as ``proposed: M=2,'' ``proposed: M=4,'' and ``proposed: M=6,'' in Fig.\ \ref{fig0}. The tuning parameters $\lambda$ and $\alpha$ were selected by searching over a grid of values to maximize the $F_1$-score (over 100 runs). The search for $\alpha$ was confined to $[0,0.3]$. For a fixed $\alpha = 0.1$, we first picked the best $\lambda$ value, and then with fixed best $\lambda$ value, search over $\alpha \in [0,0.3]$. Fig.\ \ref{fig0} shows the results for thus optimized $(\lambda, \alpha)$. In practice, one cannot calculate the $F_1$-score since ground truth is unknown. For $M=4$ we selected  $(\lambda, \alpha)$ in each run via BIC as discussed in Sec.\ \ref{BIC} (where knowledge of the ground truth is not needed). The obtained results based on 50 runs are shown in Fig.\ \ref{fig0}, labeled as ``proposed: M=4,BIC.'' The conventional i.i.d.\ modeling approach exploits only the sample covariance $\frac{1}{n} \sum_{t=0}^{n-1} {\bm x}(t) {\bm x}^\top(t)$ whereas the proposed approaches exploit the entire correlation function (equivalently PSD), and thus, can deliver better performance. In Fig.\ \ref{fig0}, the label ``IID model'' stands for the ADMM lasso approach (\cite[Sec.\ 6.4]{Boyd2010}) that models data as i.i.d., and the corresponding results are based on 100 runs with lasso parameter $\lambda$ selected by exhaustive search over a grid of values to maximize $F_1$ score. We also show the results of the ADMM approach of \cite{Jung2015a}, labeled ``GMS'' (graphical model selection), which was applied with $F=4$ (four frequency points, corresponds to our $M=4$) and all other default settings of \cite{Jung2015a} to compute the PSDs (see also Sec.\ \ref{compare}).  The lasso parameter $\lambda$ for \cite{Jung2015a} was selected by exhaustive search over a grid of values to maximize $F_1$ score. 

\begin{figure}[h]
  \centering
  \includegraphics[width=1.0\linewidth]{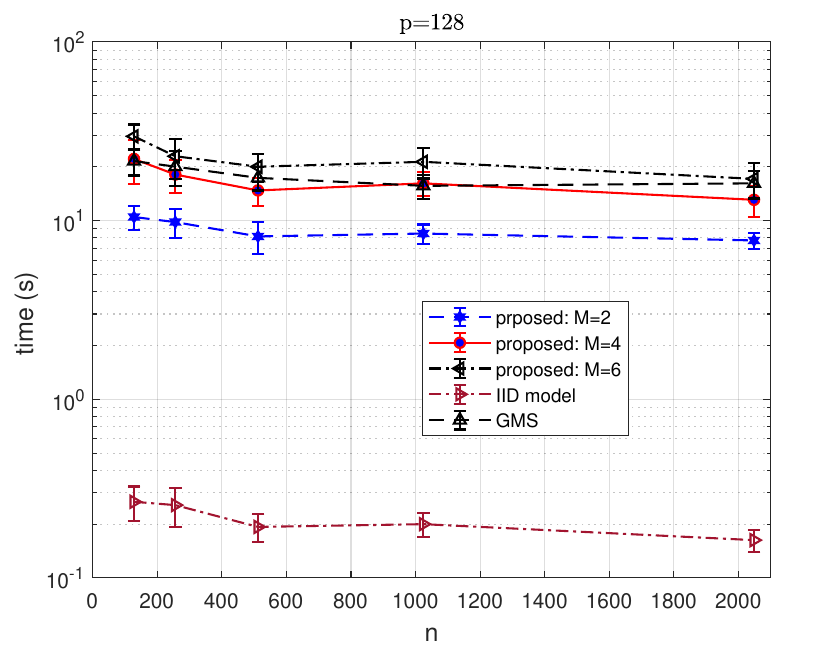} 
\caption{Average timing per run for synthetic data example. The label ``GMS'' refers to the approach of \cite{Jung2015a}.}
\label{fig0b}
\end{figure}

The $F_1$-scores are shown in Fig.\ \ref{fig0} and average timings per run are shown in Fig.\ \ref{fig0b} for sample sizes $n=128,256,512,1024, 2048$. All ADMM algorithms were implemented in MATLAB R2020b, and run on a Window Home 10 operating system with processor Intel(R) Core(TM) i5-6400T CPU @2.20 GHz with 12 GB RAM.  It is seen from Fig.\ \ref{fig0} that with $F_1$-score as the performance metric, our proposed approach for all three values of $M$ (number of normalized frequency points in $(0,0.5)$) significantly outperforms the conventional IID model approach. The BIC-based approach for $M=4$ yields performance that is close to that based on optimized parameter selection for $n \ge 512$.  It is also seen that \cite{Jung2015a} performs better than IID modeling but much worse than our proposed sparse group lasso approach, while also taking more time to convergence. The stopping rule, variable penalty and thresholds selected for all approaches were the same since all approaches are ADMM-based; these values have been specified as $\rho^{(0)} =2$, $\mu =10$, and $\tau_{abs}=\tau_{rel} =10^{-4}$ in Sec.\ \ref{stopping}. 

The conventional i.i.d.\ modeling approach estimates the (sparse) precision matrix ${\bm \Omega} = \left( E\{ {\bm x}(t) {\bm x}^\top(t) \} \right)^{-1} = {\bm R}_{xx}^{-1}(0)$: there is an edge $\{ i,j \}$ in CIG iff $\Omega_{ij} \ne 0$. Since this approach ignores ${\bm R}_{xx}(m)$ for $m \ne 0$ for dependent data, its performance is the worst for all sample sizes, although the performance does improve with $n$ since $\{ i,j \} \not\in {\cal E} \Rightarrow \Omega_{ij} = 0$ and accuracy of the estimate of $\Omega_{ij}$ improves with increasing $n$. The method of \cite{Jung2015a} performs better than IID modeling since it does use ${\bm R}_{xx}(m)$ for $m \ne 0$ in estimating ${\bm S}_{xx}^{-1}(f)$. Also, the performance of \cite{Jung2015a} improves with $n$ as accuracy of the estimates of the PSD improves with $n$. However, as summarized in Sec.\ \ref{compare}, \cite{Jung2015a} makes some peculiar choices which are likely reasons why its performance is inferior to our proposed approach. 

Note that for our example, there is no explicit mathematical expression for calculating the true edge set ${\cal E}_0$. In each run, we calculated the true PSD ${\bm S}(f)$ for $f \in [0,0.5]$ at intervals of 0.01, and then took $\{i,j\} \in {\cal E}$ if $\sum_f | S_{ij}^{-1}(f) | > 10^{-6}$, where the average value of diagonal elements $(\sum_{i=1}^p \sum_f S_{ii}^{-1}(f))/p$, averaged over 100 runs, turns out to be 75.10 ($\pm 1.85$). That is, we have some very ``weak'' edges in the graph which are not easy to detect with relatively ``short'' sample sizes, resulting in relatively low $F_1$ scores. Nevertheless, when comparing different approaches, our proposed approach performs much better.

\begin{figure}[h]
\begin{subfigure}{.5\textwidth}
  \centering
  \includegraphics[width=.8\linewidth]{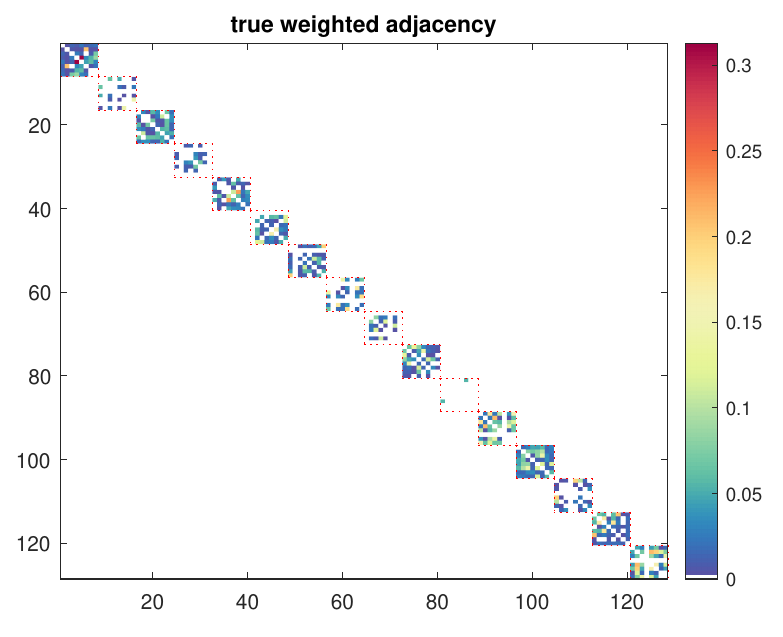} 
  \caption{True $|\Omega_{ij}|$ as edge weight.}
  \label{fig1a}
\end{subfigure}
\begin{subfigure}{.5\textwidth}
  \centering
  \includegraphics[width=.8\linewidth]{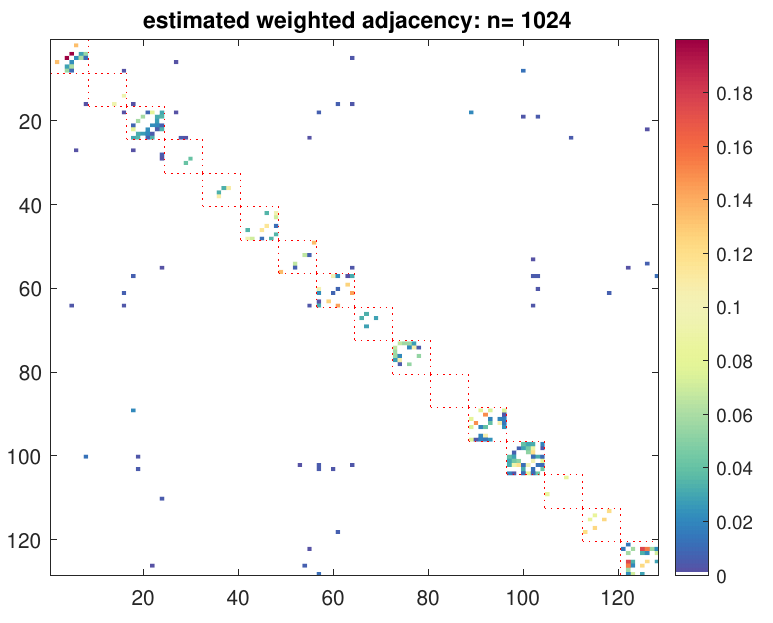} 
  \caption{Estimated $|\Omega_{ij}|$ as edge weight.}
  \label{fig1b}
\end{subfigure}
\caption{IID modeling-based weighted adjacency matrices. The red squares (in dotted lines) show the communities -- they are not part of the adjacency matrices.}
\label{fig1}
\end{figure}

\begin{figure}[h]
\begin{subfigure}{.5\textwidth}
  \centering
  \includegraphics[width=.8\linewidth]{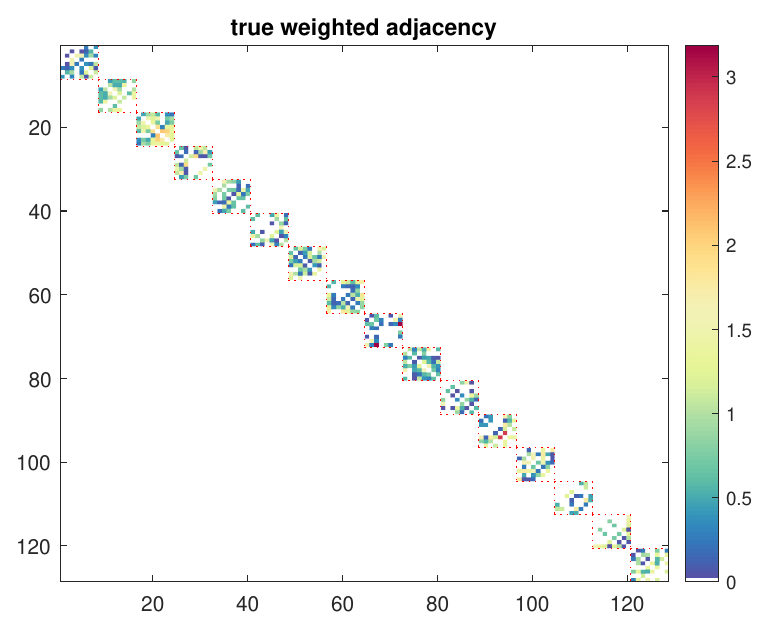} 
  \caption{True $\sqrt{\sum_{k=1}^M | [{\bm{\Phi}}_k]_{ij} |^2 }$ as edge weight.}
  \label{fig2a}
\end{subfigure}
\begin{subfigure}{.5\textwidth}
  \centering
  \includegraphics[width=.8\linewidth]{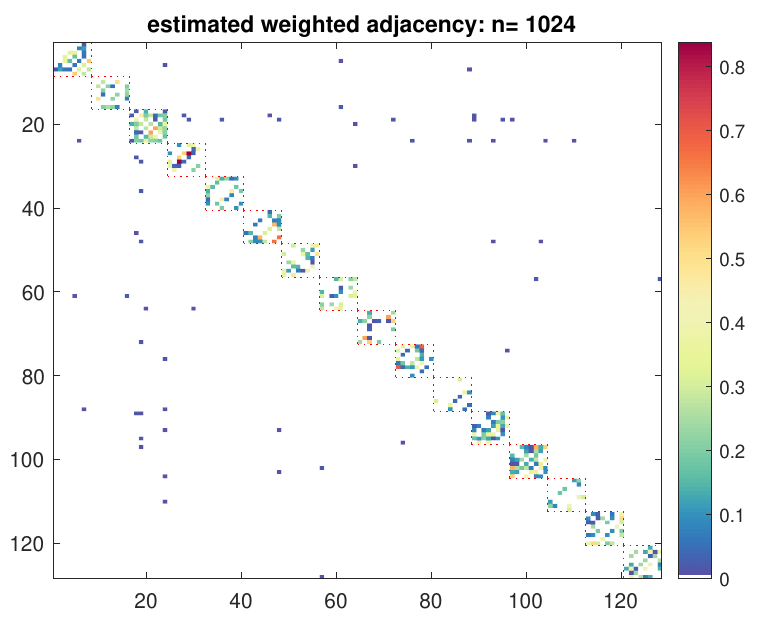} 
  \caption{Estimated $\sqrt{\sum_{k=1}^M | [\hat{\bm{\Phi}}_k]_{ij} |^2 }$ as edge weight.}
  \label{fig2b}
\end{subfigure}
\caption{Weighted adjacency matrices for dependent time series modeling: $M=4$. The red squares (in dotted lines) show the communities -- they are not part of the adjacency matrices.}
\label{fig2}
\end{figure}

For a typical Monte Carlo run with $n=1024$, we show the estimated weighted adjacency matrices resulting from the conventional ``IID model'' approach and from the ``proposed: M=4,BIC'' approach in Figs.\ \ref{fig1} and \ref{fig2} respectively. For the IID model approach, tuning parameter $\lambda$ used is the one used for Fig.\ \ref{fig0}, selected by exhaustive search to maximize the $F_1$ score.
Fig.\ \ref{fig1} shows true and estimated $|\Omega_{ij}|$ as edge weights, whereas Fig.\ \ref{fig2} shows true $\sqrt{\sum_{k=1}^M | [{\bm{\Phi}}_k]_{ij} |^2 }$ and estimated $\sqrt{\sum_{k=1}^M | [\hat{\bm{\Phi}}_k]_{ij} |^2 }$ as edge weights. While clustering is quite evident in both Figs.\ \ref{fig1} and \ref{fig2}, there are some spurious (as well as missed) connections reflecting estimation errors, which are inevitable for any finite sample size.

\subsection{Real data: Financial Time Series} 
\begin{figure}[ht]
\begin{subfigure}{.5\textwidth}
  \centering
  \includegraphics[width=.8\linewidth]{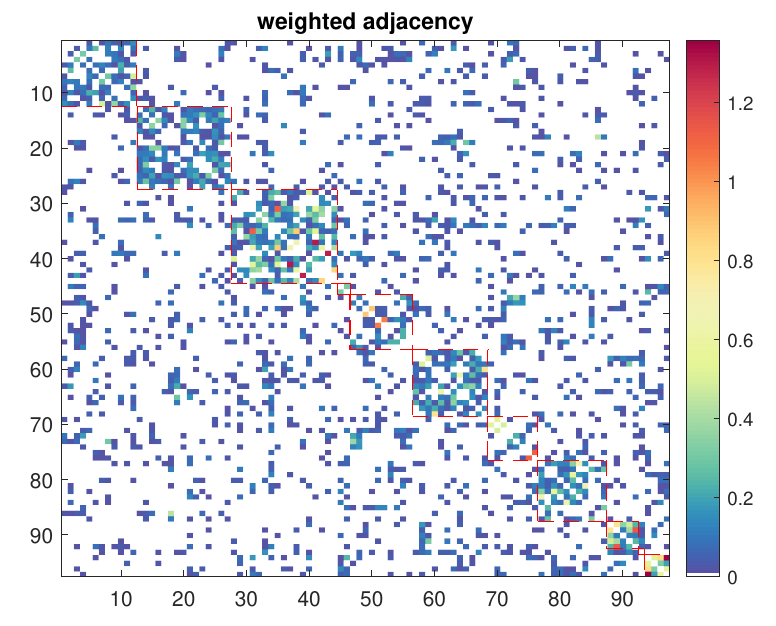} 
  \caption{Estimated $|\Omega_{ij}|$ as edge weight; 1293 edges.}
  \label{fig3a}
\end{subfigure}
\begin{subfigure}{.5\textwidth}
  \centering
  \includegraphics[width=.8\linewidth]{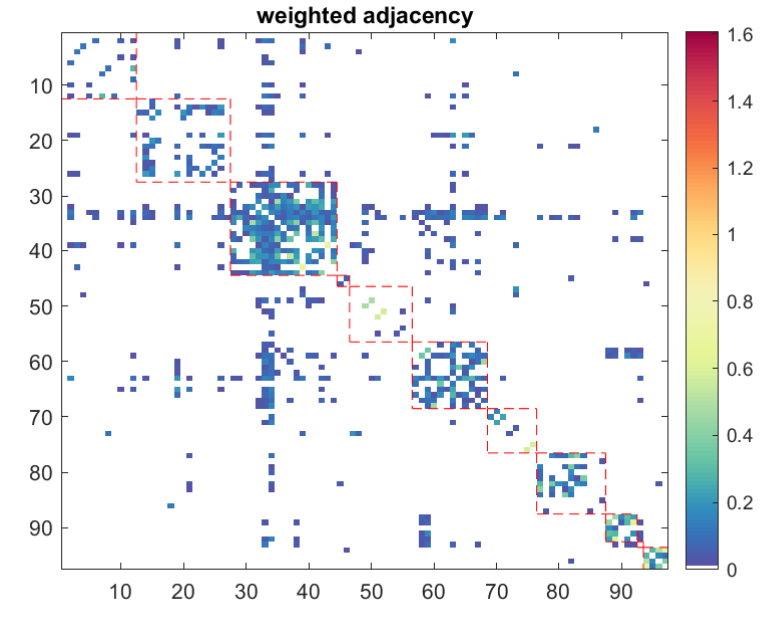} 
  \caption{Estimated $\sqrt{\sum_{k=1}^M | [\hat{\bm{\Phi}}_k]_{ij} |^2 }$ as edge weight; 429 edges.}
  \label{fig3b}
\end{subfigure}
\caption{Weighted adjacency matrices for financial time series, $p=97$, $n=1258$. (a) IID model approach, (b) Proposed approach, $M=4$. The red squares (in dashed lines) show the 11 GISC sectors -- they are not part of the adjacency matrices.}
\label{fig3}
\end{figure}

We consider daily share prices (at close of the day) of 97 stocks in S\&P 100 index from Jan. 1, 2013 through Jan.\ 1, 2018, yielding 1259 samples. This data was gathered from Yahoo Finance website. If $y_m(t)$ is the share price of $m$th stock on day $t$, we consider (as is conventional in such studies \cite{Abdelwahab2008}) $x_m(t) = \ln (y_m(t)/y_m(t-1))$ as the time series to analyze, yielding $n=1258$ and $p=97$. These 97 stocks are classified into 11 sectors (according to the Global Industry Classification Standard (GICS)), and we order the nodes to group them according to GICS sectors as information technology (nodes 1-12), health care (13-27), financials (28-44), real estate (45-46), consumer discretionary (47-56), industrials (57-68), communication services (69-76), consumer staples (77-87), energy (88-92), materials (93), and utilities (94-97). The weighted adjacency matrices resulting from the conventional i.i.d.\ modeling approach and the proposed approach with $M=4$, ($K=155$), are shown in Fig.\ \ref{fig3}. In both cases we used BIC to determine the tuning parameters with selected $\lambda = 0.0387$ for the IID model and $(\lambda , \alpha) = (0.7, 0.3)$ for the proposed approach. While the ground truth is unknown, the dependent time series based proposed approach yields sparser CIG (429 edges for the proposed approach versus 1293 edges for conventional modeling, where we now count edges $\{i,j\}$ and $\{j,i\}$, $i\ne j$, as the same one edge). Based on the GICS sector classification, one expects to see clustering in the estimated weighted adjacency matrix, conforming to the GICS classification in that stocks within a given sector are more connected, and with higher weights, to other stocks within the sector, and have fewer connections, and with lower weights, to stocks in other sectors.  In this sense, our proposed approach also conforms better with the GICS sector classification when compared to the i.i.d.\ modeling approach.

\section{Conclusions} Graphical modeling of dependent Gaussian time series was considered. A sparse-group lasso-based frequency-domain formulation of the problem was proposed and analyzed where the objective was to estimate the inverse PSD of the data via optimization of a sparse-group lasso penalized log-likelihood cost function. The graphical model is then inferred from the estimated inverse PSD. We investigated an ADMM approach for optimization. We established sufficient conditions for convergence in the Frobenius norm of the inverse PSD estimators to the true value, jointly across all frequencies, where the number of frequencies were allowed to increase with sample size. We also empirically investigated selection of the tuning parameters based on the Bayesian information criterion, and illustrated our approach using numerical examples utilizing both synthetic and real data. The synthetic data results show that for graph edge detection, the proposed approach significantly outperformed the widely used i.i.d.\ modeling approach where the underlying time series is either assumed to be i.i.d., or one uses only the covariance of the data. The proposed approach also outperformed the approach of \cite{Jung2015a} for synthetic data.

\appendices

\section{Proof of Lemma 1} \label{proofL1}
Lemma 1 is proved following \cite{Friedman2010a} (which deals with real variables), and using Wirtinger calculus and complex subgradients. Since $h(\bm{\theta})$ is convex in $\bm{\theta}$, a necessary and sufficient condition for a global minimum at $\hat{\bm{\theta}}$ is that the subdifferential of $h(\bm{\theta})$ at $\hat{\bm{\theta}}$, $\partial h(\hat{\bm{\theta}})$ given by (\ref{eqlem1_2010}), must contain ${\bm 0}$:
\begin{equation}  \label{eqlem1_2010}
   {\bm 0} \in \partial h(\hat{\bm{\theta}}) = \frac{1}{2} (\hat{\bm{\theta}} - {\bm a}) 
	    + \frac{\lambda_1}{2} {\bm t} + \frac{\lambda_2}{2} {\bm w}
\end{equation} 
where ($t_j = [{\bm t}]_j$, $j$th component of ${\bm t}$)
\begin{equation}  \label{eqlem1_2010b}
  {\bm w} = \left\{ \begin{array}{ll}
			    \hat{\bm{\theta}}/\|\hat{\bm{\theta}}\| & \mbox{if } \hat{\bm{\theta}} \neq {\bm 0} \\
					\in \{ {\bm u} \,:\, \|{\bm u}\| \le 1, \; {\bm u} \in \mathbb{C}^q \} & 
					           \mbox{if } \hat{\bm{\theta}} = {\bm 0} \end{array} \right. , 
\end{equation}
\begin{equation}  \label{eqlem1_2010c}
  t_j = \left\{ \begin{array}{ll}
			    \hat{\theta}_j/|\hat{\theta}_j| & \mbox{if } \hat{\theta}_j \neq 0 \\
					\in \{ v \,:\, |v| \le 1, \; v \in \mathbb{C} \} & \mbox{if } \hat{\theta}_j = 0 \end{array} \right. .
\end{equation}
Lemma 1 is a consequence of KKT conditions (\ref{eqlem1_2010}). We will show that (\ref{lemsol}) satisfies (\ref{eqlem1_2010}). Consider the following two cases:
\begin{itemize}
\item[(i)] Suppose $\|{\bm S}({\bm a}, \lambda_1)\| \le \lambda_2$. Then (\ref{lemsol}) implies that $\hat{\theta}_i = 0$ $\forall i$. We need to show that this solution satisfies (\ref{eqlem1_2010}), that is, given ${\bm a}$, there exist ${\bm w}$ and ${\bm t}$ satisfying
\begin{equation}  
   w_j = \frac{a_j-\lambda_1 t_j}{\lambda_2} \, , \;\; j=1,2, \cdots , q \, .
\end{equation} 
Following real-valued results of \cite{Friedman2010a}, consider minimization of $J({\bm t})$, defined below, w.r.t.\ $t_j$'s subject to $|t_j|^2 \le 1$, $j=1,2, \cdots , q \,$:
\begin{equation}  
   J({\bm t}) = \sum_{j=1}^q |w_j|^2 = \frac{1}{\lambda_2^2} \sum_{j=1}^q  |a_j-\lambda_1 t_j|^2 \, .
\end{equation}
The problem is separable in $t_j$'s with the solution
\begin{equation}  
   \hat{t}_j = \left\{ \begin{array}{ll} 
	          a_j/\lambda_1 & \mbox{if  } |a_j| \le \lambda_1 \\
						a_j/ |a_j| & \mbox{if  } |a_j| > \lambda_1 \, . \end{array} \right.
\end{equation}
Thus $a_j - \lambda_1 \hat{t}_j = (1-\lambda_1/|a_j|)_+ a_j$ and with $\hat{w}_j = (a_j - \lambda_1 \hat{t}_j)/\lambda_2$, we have
\begin{equation}  
   \min_{{\bm t}} J({\bm t}) = J(\hat{\bm t}) = \sum_{j=1}^q |\hat{w}_j|^2 
	 = \frac{1}{\lambda_2^2} \|{\bm S}({\bm a}, \lambda_1)\|^2 \le 1 \, .
\end{equation}
Thus (\ref{eqlem1_2010}) holds for given $\hat{\bm w}$ and $\hat{\bm t}$ satisfying (\ref{eqlem1_2010b}) and (\ref{eqlem1_2010c}), respectively.
\item[(ii)] Now suppose $\|{\bm S}({\bm a}, \lambda_1)\| > \lambda_2$. Then (\ref{lemsol}) implies that $\hat{\bm{\theta}} \ne {\bm 0}$, therefore, at least one component of $\hat{\bm{\theta}}$ is nonzero. Again we need to show that this solution satisfies (\ref{eqlem1_2010}). If $|a_i| \le \lambda_1$, then $\hat{\theta}_i = 0$ and $\hat{t}_i = a_i/\lambda_1$ satisfies (\ref{eqlem1_2010c}). If $|a_i| > \lambda_1$, then $\hat{\theta}_i \ne 0$ and we take $\hat{t}_i = \hat{\theta}_i/|\hat{\theta}_i|$. For an arbitrary $a_i$, but satisfying $\|{\bm S}({\bm a}, \lambda_1)\| > \lambda_2$, let us express our claimed solution $\hat{\theta}_i \ne 0$ as 
\begin{align}  
   \hat{\theta}_i = &\gamma \alpha_i a_i \, , \quad \alpha_i := (1-\lambda_1/|a_i|)_+  \\
	  \gamma = & 1- \lambda_2/(\sum_{j=1}^q \alpha_j^2 |a_j|^2 )^{1/2}  > 0 \, .
\end{align}
We need to show that this solution satisfies (\ref{eqlem1_2010}). The $i$th component of $2 \partial h(\hat{\bm{\theta}})$ with $\hat{\bm w}$ and $\hat{\bm t}$ satisfying (\ref{eqlem1_2010b}) and (\ref{eqlem1_2010c}), respectively, is
\begin{align}  
   A := & \hat{\theta}_i - a_i + \lambda_1 \frac{\hat{\theta}_i}{|\hat{\theta}_i|} 
	          + \lambda_2 \frac{\hat{\theta}_i}{\|\hat{\bm \theta}\|}  \nonumber \\
			= & \hat{\theta}_i \left[ 1 + \frac{\lambda_1}{|\hat{\theta}_i|} 
			       +  \frac{\lambda_2}{\|\hat{\bm \theta}\|} \right] - a_i = a_i B - a_i
\end{align}
where, with $D := (\sum_{j=1}^q \alpha_j^2 |a_j|^2)^{1/2}$,
\begin{align}  
   B = &\gamma \alpha_i + \frac{\lambda_1}{|a_i|} 
			       +  \frac{\lambda_2 \alpha_i}{D} \, .
\end{align} 
The proof is completed by showing that $B=1$. We have
\begin{align}
  B = & \frac{ \gamma \alpha_i |a_i| D + \lambda_1 D +  \lambda_2 \alpha_i |a_i|}{ |a_i| D }  \\
	= & \frac{ \alpha_i |a_i| D - \lambda_2 \alpha_i |a_i|+ \lambda_1 D + \lambda_2 \alpha_i |a_i|}{ |a_i| D }  \\
			= & \frac{\alpha_i |a_i| + \lambda_1}{|a_i|} = \alpha_1 + \frac{\lambda_1}{|a_i|} = 1 
\end{align}
where, for $\hat{\theta}_i \ne 0$, $\alpha_i = 1-\lambda_1/|a_i|$.  
\end{itemize}
This proves the desired result $\quad \blacksquare$ 

\section{Proof of Theorem 1} \label{proof}
Our proof relies on the method of \cite{Rothman2008} which deals with i.i.d.\ time series models and lasso penalty, and our prior results in \cite{Tugnait19c} dealing with complex Gaussian vectors (not time series). From now on we use the term ``with high probability'' (w.h.p.) to denote with probability greater than $1-1/p_n ^{\tau-2}$. First we need several auxiliary results.

Lemma 2 below is specialization of \cite[Lemma 1]{Ravikumar2011} to Gaussian random vectors. It follows from \cite[Lemma 1]{Ravikumar2011} after setting the sub-Gaussian parameter $\sigma$ in \cite[Lemma 1]{Ravikumar2011} to 1. \\
{\it Lemma 2}. Consider a zero-mean Gaussian random vector ${\bm z} \in \mathbb{R}^p$ with covariance ${\bm R} \succ {\bm 0}$. Given $n$ i.i.d.\ samples ${\bm z}(t)$, $t=0, 1, \cdots , n-1$, of ${\bf z}$, let $\hat{\bm R} =  (1/n) \sum_{t=0}^{n-1} {\bm z} {\bm z}^\top$ denote the sample covariance matrix. Then $\hat{\bm R}$ satisfies the tail bound
\begin{equation}  \label{naeq57}
   P \left( \Big| [ \hat{\bm{R}} - \bm{R} ]_{ij} \Big| > \delta \right)
	          \le 4 \, \exp  \left( - \frac{n \delta^2 }{
						  3200 \max_i ( R_{ii}^2)} \right) 
\end{equation} 
for all $\delta \in (0, 40 \max_i (R_{ii}) )$     $\quad \bullet$

Exploiting Lemma 2, we have Lemma 3 regarding $\hat{\bm S}_k$. We denote ${\bm S}_0(\tilde{f}_k)$ as ${\bm S}_{0k}$ in this section. A proof of Lemma 3 is in \cite[Lemma 2]{Tugnait20}. The statement of Lemma 3 below is a corrected version of \cite[Lemma 2]{Tugnait20} with no changes to its proof given therein. \\
{\it Lemma 3}. Under Assumption (A1), $\hat{\bm S}_k$ satisfies the tail bound
\begin{align}  
   P \Big(  \max_{k,q,l} &\Big| [ \hat{\bm S}_k - {\bm S}_{0k} ]_{ql} \Big| 
	    > C_0 \sqrt{\frac{\ln(p_n )}{K_n}} \Big)  \le \frac{1}{p_n ^{\tau -2}} \label{naeq58b}
\end{align}
for $\tau > 2$, if the sample size $n$ and choice of $K_n$ is such that $K_n >  N_1 = 2 \ln (16 p_n^\tau M_n)$, where $C_0$ is defined in (\ref{naeq58}). $\quad \bullet$ \\

Lemma 4 deals with a Taylor series expansion using Wirtinger calculus. \\
{\em Lemma 4}. For $\bm{\Phi}_k = \bm{\Phi}_k^H \succ {\bm 0}$, define a real scalar function 
\begin{equation}  \label{naeq501}
   c(\bm{\Phi}_k, \bm{\Phi}_k^\ast) = \ln |\bm{\Phi}_k| + \ln |\bm{\Phi}_k^\ast| \, .
\end{equation}
Let $\bm{\Phi}_k = \bm{\Phi}_{0k} + \bm{\Gamma}_k$ with $\bm{\Phi}_{0k} = \bm{\Phi}_{0k}^H \succ {\bm 0}$ and $\bm{\Gamma}_k = \bm{\Gamma}_k^H$. Then using Wirtinger calculus, the Taylor series expansion of $c(\bm{\Phi}_k, \bm{\Phi}_k^\ast)$ is given by
\begin{align}
   c& (\bm{\Phi}_k, \bm{\Phi}_k^\ast) =  c(\bm{\Phi}_{0k}, \bm{\Phi}_{0k}^\ast) 
	   +\mbox{tr}(\bm{\Phi}_{0k}^{-1} \bm{\Gamma}_k + \bm{\bm{\Phi}}_{0k}^{-\ast} \bm{\Gamma}_k^\ast ) \nonumber \\
		 & \;\; -\frac{1}{2} (\mbox{vec}(\bm{\Gamma}_k))^H ( \bm{\Phi}_{0k}^{-\ast} \otimes \bm{\Phi}_{0k}^{-1} ) 
		              \mbox{vec}(\bm{\Gamma}_k) \nonumber \\
		 & \;\; -\frac{1}{2} (\mbox{vec}(\bm{\Gamma}_k^\ast))^H ( \bm{\Phi}_{0k}^{-1} \otimes \bm{\Phi}_{0k}^{-\ast} ) 
		              \mbox{vec}(\bm{\Gamma}_k^\ast) + \mbox{h.o.t.}  \label{naeq60}
\end{align}
where h.o.t.\ stands for higher-order terms in $\bm{\Gamma}_k$ and $\bm{\Gamma}_k^\ast$. $\quad \bullet$

{\em Proof}: Only for the proof of this lemma, we will drop the subscript $k$, and donate $\bm{\Phi}_k$, $\bm{\Phi}_{0k}$ and $\bm{\Gamma}_k$ as $\bm{\Phi}$, $\bm{\Phi}_{0}$ and $\bm{\Gamma}$, respectively. Treating $\bm{\Phi}$ and $\bm{\Phi}^\ast$ as independent variables, the Taylor series expansion of $c(\bm{\Phi}, \bm{\Phi}^\ast)$ is
\begin{align}
   c & (\bm{\Phi}, \bm{\Phi}^\ast) =  c(\bm{\Phi}_0, \bm{\Phi}_0^\ast) 
	   + \sum_{s,t} \left( \frac{\partial c}{\partial \Phi_{0st}} \Gamma_{st} +
		   \frac{\partial c}{\partial \Phi_{0st}^\ast} \Gamma_{st}^\ast  \right) \nonumber \\
		& \;\; + \frac{1}{2} \sum_{q,r} \sum_{s,t} [ \Gamma_{qr} \quad \Gamma_{qr}^\ast ]
		    D_{0 qrst} \left[ \begin{array}{c}  \Gamma_{st} \\ \Gamma_{st}^\ast \end{array} \right] + \mbox{h.o.t.}
				 \label{naeq70}
\end{align}
where 
\begin{equation}
   \frac{\partial c}{\partial \Phi_{0st}} := 
	         \left. \frac{\partial c(\bm{\Phi}, \bm{\Phi}^\ast)}{\partial \Phi_{st}} 
						        \right|_{\substack{\Phi_{st} = \Phi_{0st} \\ \Phi_{st}^\ast = \Phi_{0st}^\ast}} \, ,
\end{equation}
\begin{equation}
   \frac{\partial c}{\partial \Phi_{0st}^\ast} := 
	         \left. \frac{\partial c(\bm{\Phi}, \bm{\Phi}^\ast)}{\partial \Phi_{st}^\ast} 
						        \right|_{\substack{\Phi_{st} = \Phi_{0st} \\ \Phi_{st}^\ast = \Phi_{0st}^\ast}} \, ,
\end{equation}
\begin{align}
   D_{0 qrst} = & \left. \left[ \begin{array}{cc}
	          \frac{\partial^2 c(\bm{\Phi}, \bm{\Phi}^\ast)}{\partial \Phi_{qr}\partial \Phi_{st}} &
						\frac{\partial^2 c(\bm{\Phi}, \bm{\Phi}^\ast)}{\partial \Phi_{qr}\partial \Phi_{st}^\ast} \\
			\frac{\partial^2 c(\bm{\Phi}, \bm{\Phi}^\ast)}{\partial \Phi_{qr}^\ast \partial \Phi_{st}} &
			\frac{\partial^2 c(\bm{\Phi}, \bm{\Phi}^\ast)}{\partial \Phi_{qr}^\ast \partial \Phi_{st}^\ast}			
				\end{array} \right]	\right|_{\substack{\Phi_{st} = \Phi_{0st}, \Phi_{st}^\ast = \Phi_{0st}^\ast
				             \\ \Phi_{qr} = \Phi_{0qr}, , \Phi_{qr}^\ast = \Phi_{0qr}^\ast}} \nonumber \\
	=: & \left[ \begin{array}{cc}
	          \frac{\partial^2 c(\bm{\Phi}, \bm{\Phi}^\ast)}{\partial \Phi_{0qr}\partial \Phi_{0st}} &
						\frac{\partial^2 c(\bm{\Phi}, \bm{\Phi}^\ast)}{\partial \Phi_{0qr}\partial \Phi_{0st}^\ast} \\
			\frac{\partial^2 c(\bm{\Phi}, \bm{\Phi}^\ast)}{\partial \Phi_{0qr}^\ast \partial \Phi_{0st}} &
			\frac{\partial^2 c(\bm{\Phi}, \bm{\Phi}^\ast)}{\partial \Phi_{0qr}^\ast \partial \Phi_{0st}^\ast}			
				\end{array} \right] \, .
\end{align}
Consider the following facts \cite{Dattorro2013, Petersen2005}
\begin{align}
  \frac{\partial \ln |\bm{\Phi}|}{\partial \Phi_{st}} = & [\bm{\Phi}^{-\top}]_{st} =[\bm{\Phi}^{-1}]_{ts} 
	    \mbox{ since } \frac{\partial \ln |\bm{\Phi}|}{\partial \bm{\Phi}} = \bm{\Phi}^{-\top} \, , \\
			\frac{\partial \ln |\bm{\Phi}^\ast|}{\partial \Phi_{st}^\ast} = & [\bm{\Phi}^{-\ast}]_{ts} \, , \\
	\frac{\partial \ln |\bm{\Phi}|}{\partial \Phi_{st}^\ast} = & 0, \quad 
	       \frac{\partial \ln |\bm{\Phi}^\ast|}{\partial \Phi_{st}} =  0 \, , \\
	\frac{\partial^2 \ln |\bm{\Phi}|}{\partial \Phi_{qr} \partial \Phi_{st}} =& 
	     \frac{\partial [\bm{\Phi}^{-1}]_{ts}}{\partial \Phi_{qr}}
			  = - [\bm{\Phi}^{-1}]_{tq} [\bm{\Phi}^{-1}]_{rs} \, , \\
	\frac{\partial^2 \ln |\bm{\Phi}^\ast|}{\partial \Phi_{qr}^\ast \partial \Phi_{st}^\ast} =& 
			   - [\bm{\Phi}^{-\ast}]_{tq} [\bm{\Phi}^{-\ast}]_{rs} \, , \\
	\frac{\partial^2 \ln |\bm{\Phi}|}{\partial \Phi_{qr} \partial \Phi_{st}^\ast} =& 
			   \frac{\partial^2 \ln |\bm{\Phi}|}{\partial \Phi_{qr}^\ast \partial \Phi_{st}} =
				 \frac{\partial^2 \ln |\bm{\Phi}^\ast| }{\partial \Phi_{qr} \partial \Phi_{st}^\ast} = 
			   \frac{\partial^2 \ln |\bm{\Phi}^\ast| }{\partial \Phi_{qr}^\ast \partial \Phi_{st}}= 0 \, . 
\end{align}

Using the above partial derivatives in the first derivative terms in the Taylor series, we have 
\begin{align}
  \sum_{s,t} & \left( \frac{\partial c}{\partial \Phi_{0st}} \Gamma_{st} +
		   \frac{\partial c}{\partial \Phi_{0st}^\ast} \Gamma_{st}^\ast \right) \nonumber \\
		& = \sum_{s,t} \left( [\bm{\Phi}_0^{-1}]_{ts} \Gamma_{st} +  [\bm{\Phi}_0^{-\ast}]_{ts} \Gamma_{st}^\ast  \right) \nonumber \\
			& = \mbox{tr}(\bm{\Phi}_0^{-1} \bm{\Gamma} + \bm{\Phi}_0^{-\ast} \bm{\Gamma}^\ast ) \, .
			  \label{naeq75}
\end{align}
The quadratic terms in the Taylor series yield
\begin{align}
  \sum_{q,r} & \sum_{s,t} [ \Gamma_{qr} \quad \Gamma_{qr}^\ast ]
		    D_{0 qrst} \left[ \begin{array}{c}  \Gamma_{st} \\ \Gamma_{st}^\ast \end{array} \right] \nonumber \\
		& = \sum_{q,r} \sum_{s,t} [ \Gamma_{qr} \frac{\partial^2 c(\bm{\Phi}, \bm{\Phi}^\ast)}{\partial \Phi_{0qr}\partial \Phi_{0st}}  \Gamma_{st} + \Gamma_{qr}^\ast \frac{\partial^2 c(\bm{\Phi}, \bm{\Phi}^\ast)}{\partial \Phi_{0qr}^\ast \partial \Phi_{0st}^\ast} \Gamma_{st}^\ast ] \nonumber \\
		& = - \sum_{q,r} \sum_{s,t} \Big[ \Gamma_{qr} [\bm{\Phi}_0^{-1}]_{tq} [\bm{\Phi}_0^{-1}]_{rs}  \Gamma_{st} \nonumber \\
		    & \quad\quad\quad\quad + \Gamma_{qr}^\ast [\bm{\Phi}_0^{-\ast}]_{tq} [\bm{\Phi}_0^{-\ast}]_{rs}  \Gamma_{st}^\ast \Big] \nonumber \\
	& = - \sum_{q,s} \left[ \left( \sum_r \Gamma_{qr} [\bm{\Phi}_0^{-1}]_{rs} \right) + \left( \sum_t \Gamma_{st} [\bm{\Phi}_0^{-1}]_{tq} \right) \right] \nonumber \\
		    & \quad\quad - \sum_{q,s} \left[ \left( \sum_r \Gamma_{qr}^\ast [\bm{\Phi}_0^{-\ast}]_{rs} \right) + \left( \sum_t \Gamma_{st}^\ast [\bm{\Phi}_0^{-\ast}]_{tq} \right)  \right] \nonumber \\
  & = - \sum_{q,s} \left[  [\bm{\Gamma}  \bm{\Phi}_0^{-1}]_{qs}  [\bm{\Gamma} \bm{\Phi}_0^{-1}]_{sq}  +
	   [\bm{\Gamma}^\ast \bm{\Phi}_0^{-\ast}]_{qs}  [\bm{\Gamma}^\ast \bm{\Phi}_0^{-\ast}]_{sq}   \right] \nonumber \\
			& = - \mbox{tr} \left( \bm{\Gamma} \bm{\Phi}_0^{-1} \bm{\Gamma} \bm{\Phi}_0^{-1}  
			     + \bm{\Gamma}^\ast \bm{\Phi}_0^{-\ast} \bm{\Gamma}^\ast \bm{\Phi}_0^{-\ast} \right) \, .
					       \label{naeq190}
\end{align}

Given matrices ${\bm A}$ and ${\bm B}$ for which product ${\bm A} {\bm B}$ is defined, and additionally given matrix ${\bm Y}$ such that product ${\bm A} {\bm Y} {\bm B}$ is defined, we have $\mbox{vec}({\bm A} {\bm Y} {\bm B}) = ({\bm B}^\top \otimes {\bm A}) \mbox{vec}({\bm Y})$ and $\mbox{tr}({\bm A} {\bm B}) = (\mbox{vec}({\bm A}))^\top \mbox{vec}({\bm B})$. Using these results we have
\begin{align}  
   (\mbox{vec}({\bm A}))^\top ({\bm D} \otimes {\bm B}) \mbox{vec}({\bm C})
	  = & (\mbox{vec}({\bm A}))^\top \mbox{vec}({\bm B} {\bm C} {\bm D}^\top) \nonumber \\
		= & \mbox{tr} \left( {\bm A}^\top {\bm B} {\bm C} {\bm D}^\top \right) \, . \label{naeq200}
\end{align}
Using (\ref{naeq200}), we rewrite terms in (\ref{naeq190}) as
\begin{align}  
   \mbox{tr} \left( \bm{\Gamma} \bm{\Phi}_0^{-1} \bm{\Gamma} \bm{\Phi}_0^{-1} \right)
	  = & (\mbox{vec}(\bm{\Gamma}^\top))^\top \left( \bm{\Phi}_0^{-\top} \otimes \bm{\Phi}_0^{-1} \right)
		         \mbox{vec}(\bm{\Gamma})  \nonumber \\
		= & (\mbox{vec}(\bm{\Gamma}))^H \left( \bm{\Phi}_0^{-\ast} \otimes \bm{\Phi}_0^{-1} \right)
		         \mbox{vec}(\bm{\Gamma}) \, , \label{naeq210} \\
		\mbox{tr} \left( \bm{\Gamma}^\ast \bm{\Phi}_0^{-\ast} \bm{\Gamma}^\ast \bm{\Phi}_0^{-\ast} \right)
	  = & (\mbox{vec}(\bm{\Gamma}^H))^\top \left( \bm{\Phi}_0^{-H} \otimes \bm{\Phi}_0^{-\ast} \right)
		         \mbox{vec}(\bm{\Gamma}^\ast) \, . \label{naeq220}
\end{align}
In (\ref{naeq210}), we have used $(\mbox{vec}(\bm{\Gamma}^\top))^\top = (\mbox{vec}(\bm{\Gamma}^\ast))^\top
= (\mbox{vec}(\bm{\Gamma}))^H$ and $\bm{\Phi}_0^{-\top} = (\bm{\Phi}_0^{\top})^{-1} = \bm{\Phi}_0^{-\ast}$, since $\bm{\Gamma} = \bm{\Gamma}^H$ and $\bm{\Phi}_0 = \bm{\Phi}_0^H$. In (\ref{naeq220}), 
we have used $(\mbox{vec}(\bm{\Gamma}^H))^\top = (\mbox{vec}(\bm{\Gamma}))^\top
= (\mbox{vec}(\bm{\Gamma}^\ast))^H$ and $\bm{\Phi}_0^{-H}  = (\bm{\Phi}_0^H)^{-1} = \bm{\Phi}_0^{-1}$. Using (\ref{naeq75}), (\ref{naeq190}), (\ref{naeq210}) and (\ref{naeq220}) in (\ref{naeq70}), we obtain the desired result (\ref{naeq60}).
$\quad \blacksquare$

Lemma 4 regarding Taylor series expansion immediately leads to Lemma 5 regarding Taylor series with integral remainder, needed to follow the proof of \cite{Rothman2008} pertaining to the  real-valued case.

{\em Lemma 5}. With $c(\bm{\Phi}_k, \bm{\Phi}_k^\ast)$ and  $\bm{\Phi}_k = \bm{\Phi}_{0k} + \bm{\Gamma}_k$ as in Lemma 4, the Taylor series expansion of $c(\bm{\Phi}_k, \bm{\Phi}_k^\ast)$ in integral remainder form is given by ($v$ is real)
\begin{align}
   c& (\bm{\Phi}_k, \bm{\Phi}_k^\ast) =  c(\bm{\Phi}_{0k}, \bm{\Phi}_{0k}^\ast) 
	   +\mbox{tr}(\bm{\Phi}_{0k}^{-1} \bm{\Gamma}_k + \bm{\bm{\Phi}}_{0k}^{-\ast} \bm{\Gamma}_k^\ast ) \nonumber \\
		  & \;\; - \bm{g}^H(\bm{\Gamma}_k) \left( \int_0^1 (1-v) 
			  \bm{H}(\bm{\Phi}_{0k}, \bm{\Gamma}_k, v ) \, dv \right)  \bm{g}(\bm{\Gamma}_k)   \label{naeq100}
\end{align}
where 
\begin{equation}
   \bm{g}(\bm{\Gamma}_k) = \left[ \begin{array}{c} \mbox{vec}(\bm{\Gamma}_k) \\
	       \mbox{vec}(\bm{\Gamma}_k^\ast)  \end{array} \right] , \;
  \bm{H}(\bm{\Phi}_{0k},\bm{\Gamma}_k, v ) 
	    = \left[ \begin{array}{cc} \bm{H}_{11k} &  \bm{0} \\
	       \bm{0} & \bm{H}_{22k}  \end{array} \right] 
\end{equation}
\begin{equation}
   \bm{H}_{11k} = (\bm{\Phi}_{0k} +v \bm{\Gamma}_k)^{-\ast} \otimes (\bm{\Phi}_{0k} +v \bm{\Gamma}_k)^{-1}
\end{equation}
and
\begin{equation}
  \bm{H}_{22k} = (\bm{\Phi}_{0k}+v \bm{\Gamma}_k)^{-1} \otimes (\bm{\Phi}_{0k}+v \bm{\Gamma}_k)^{-\ast}  \quad \bullet
\end{equation}

We now turn to the proof of Theorem 1. \\
{\em Proof of Theorem 1}. Let $\bm{\Omega} = \bm{\Omega}_0 + \bm{\Delta}$ where
\begin{align}
   \bm{\Delta} & = \left[ \bm{\Gamma}_1 \; \bm{\Gamma}_2 \; \cdots \; \bm{\Gamma}_{M_n} \right] \\
	 \bm{\Gamma}_k & = \bm{\Phi}_k - \bm{\Phi}_{0k}, \; k=1,2, \cdots , M_n ,
\end{align}
and $\bm{\Phi}_k$, $\bm{\Phi}_{0k}$ are both Hermitian positive-definite, implying $\bm{\Gamma}_k = \bm{\Gamma}_k^H$.
Let
\begin{equation}
  Q(\bm{\Omega}) := {\cal L}_{SGL}(\bm{\Omega}) 
	           - {\cal L}_{SGL}(\bm{\Omega}_0 ) \, .
\end{equation}
The estimate $\hat{\bm{\Omega}}_{\lambda}$, denoted by $\hat{\bm{\Omega}}$ hereafter suppressing dependence upon $\lambda$, minimizes $Q(\bm{\Omega})$, or equivalently, $\hat{\bm{\Delta}} = \hat{\bm{\Omega}} - \bm{\Omega}_0$ minimizes $G(\bm{\Delta}) := Q(\bm{\Omega}_0 + \bm{\Delta})$.
We will follow the method of proof of \cite[Theorem 1]{Rothman2008} pertaining to real-valued i.i.d.\ time series. Consider the set
\begin{equation}  \label{naeq1001}
  \Theta_n(R) :=  \left\{ \bm{\Delta} \, :\, \bm{\Gamma}_k = \bm{\Gamma}_k^H \; \forall k, \; \|\bm{\Delta} \|_F = R r_n \right\}
\end{equation}
where $R$ and $r_n$ are as in (\ref{neq15ab0}) and (\ref{neq15ab1}), respectively. Observe that $G(\bm{\Delta})$ is a convex function of $\bm{\Delta}$, and 
\begin{equation}
     G(\hat{\bm{\Delta}}) = Q(\bm{\Omega}_0 + \hat{\bm{\Delta}}) \le G(\bm{0}) = 0 \, .
\end{equation} 
Therefore, if we can show that 
\begin{equation}
     \inf_{\bm{\Delta}}  \{ G(\bm{\Delta}) \, :\, \bm{\Delta} \in \Theta_n(R) \} \, > \, 0 \, ,
\end{equation}
the minimizer $\hat{\bm{\Delta}}$ must be inside the sphere defined by $\Theta_n(R)$, and hence
\begin{equation} \label{more10}
     \| \hat{\bm{\Delta}} \|_F \le   R r_n \, .
\end{equation}

Using Lemma 5 we rewrite $G({\bm{\Delta}})$ as 
\begin{equation}
     G({\bm{\Delta}}) = \sum_{k=1}^{M_n} ( \frac{1}{2}A_{1k} + \frac{1}{2} A_{2k} + A_{3k}) + A_4 \, , \label{Gmain}
\end{equation}
where, noting that $\bm{\Phi}_{0k}^{-1} = \bm{S}_{0k}$, 
\begin{align} 
     A_{1k} = & \bm{g}^H(\bm{\Gamma}_k) \left( \int_0^1 (1-v) 
			  \bm{H}(\bm{\Phi}_{0k}, \bm{\Gamma}_k, v ) \, dv \right)  \bm{g}(\bm{\Gamma}_k)  \, ,  \label{naeq1100} \\
     A_{2k} = &  \mbox{tr} \left( (\hat{\bm{S}}_k - \bm{S}_{0k} ) \bm{\Gamma}_k  + 
		  (\hat{\bm{S}}_k - \bm{S}_{0k} )^\ast \bm{\Gamma}_k^\ast \right) \, , \label{naeq1110} \\
     A_{3k} = & \alpha \lambda_n ( \| \bm{\Phi}_{0k}^{-} + \bm{\Gamma}_k^{-} \|_1 
		        - \| \bm{\Phi}_{0k}^{-} \|_1 )\, , \label{naeq1120}  \\
		A_{4} = & (1-\alpha) \lambda_n \sum_{i\ne j}^{p_n} ( \| \bm{\Omega}_0^{(ij)} + \bm{\Delta}^{(ij)} \|_F 
		         - \| \bm{\Omega}_0^{(ij)} \|_F )\, , \label{naeq1122} \\
     \bm{\Omega}_0^{(ij)} := & [ [{\bm \Phi}_{01}]_{ij} \;   \cdots \; [{\bm \Phi}_{0M_n}]_{ij} ]^\top
			  \in \mathbb{C}^{M_n} \, ,  \label{naeq1100a} \\
     \bm{\Delta}^{(ij)} := & [ [{\bm \Gamma}_{1}]_{ij} \;   \cdots \; [{\bm \Gamma}_{M_n}]_{ij} ]^\top
			  \in \mathbb{C}^{M_n} \, . \label{naeq1110b} 
\end{align}

Define
\begin{equation}
  d_{1n} := \sqrt{\frac{\ln(p_n ) }{ K_n} }, \;\;
	d_{2n} : = d_{1n} \sqrt{ (p_n+s_{n0})  } \, .
\end{equation}
We first bound $A_{1k}$'s and $A_1$. Note that $\bm{H}(\bm{\Phi}_{0k}, \bm{\Gamma}_k, v )$ is a Hermitian matrix and its eigenvalues consist of the eigenvalues of $\bm{H}_{11k}$ and $\bm{H}_{22k}$. Since the eigenvalues of ${\bm A} \otimes {\bm B}$ are the product of the eigenvalues of ${\bm A}$ and eigenvalues of  ${\bm B}$ for Hermitian  ${\bm A}$ and ${\bm B}$, the eigenvalues of $\bm{H}_{11k}$ are the product of the eigenvalues of $(\bm{\Phi}_{0k}+v\bm{\Gamma}_k)^{-\ast}$ and that of $(\bm{\Phi}_{0k}+v\bm{\Gamma}_k)^{-1}$. But these two matrices have the same set of eigenvalues since one matrix is the complex conjugate of the other, and both have real eigenvalues since they are Hermitian. Since $\bm{H}_{11k} = \bm{H}_{22k}^\ast$, it follows that the eigenvalues of $\bm{H}_{11k}$ are the same as the eigenvalues of $\bm{H}_{22k}$. Thus
\begin{align}
     \phi_{\min } & (\bm{H}(\bm{\Phi}_{0k}, \bm{\Gamma}_k, v )  )  
		   = \phi_{\min }(\bm{H}_{11k}) = \phi_{\min }(\bm{H}_{22k})  \nonumber \\
			&  = \phi_{\min }^2((\bm{\Phi}_{0k}+v\bm{\Gamma}_k)^{-1}) 
			   = \phi_{\max }^{-2}(\bm{\Phi}_{0k}+v\bm{\Gamma}_k) \, .
\end{align}
Since ${\bm x}^H {\bm A} {\bm x} \ge \phi_{\min }({\bm A}) \|{\bm x}\|^2$, we have
\begin{align}
    & A_{1k}  \ge \| \bm{g}(\bm{\Gamma}_k) \|^2 \phi_{\min } \left( \int_0^1 (1-v) 
			  \bm{H}(\bm{\Phi}_{0k}, \bm{\Gamma}_k, v ) \, dv \right)  \nonumber \\
			& \; \ge 2 \|\mbox{vec}(\bm{\Gamma}_k)\|^2  \int_0^1 (1-v) \, dv \min_{0 \le v \le 1} 
			        \phi_{\min } (\bm{H}(\bm{\Phi}_{0k}, \bm{\Gamma}_k, v ))  \nonumber \\
			& \; =  \| \bm{\Gamma}_k \|_F^2  \min_{0 \le v \le 1} \phi_{\max }^{-2}(\bm{\Phi}_{0k}+v\bm{\Gamma}_k) \, ,
\end{align}
where we have used the facts that $\int_0^1 (1-v) \, dv = 1/2$. Since
\begin{equation}
     \phi_{\max }(\bm{\Phi}_{0k}+v \bm{\Gamma}_k) \le \| \bm{\Phi}_{0k}+v \bm{\Gamma}_k \|
		    \le \| \bm{\Phi}_{0k} \| + v \| \bm{\Gamma}_k \| \, ,
\end{equation}
we have
\begin{align}
     \phi_{\max }^{-2}(\bm{\Phi}_{0k}+v\bm{\Gamma}_k) & \ge (\| \bm{\Phi}_{0k} \| + v \| \bm{\Gamma}_k \| )^{-2} \nonumber \\
		   & \ge (\| \bm{\Phi}_{0k} \| +  \| \bm{\Gamma}_k \| )^{-2} \mbox{ for } 0 \le v \le 1 \, .
\end{align}
Thus,
\begin{equation}
     A_{1k}  \ge \frac{ \| \bm{\Gamma}_k \|_F^2 }{ (\| \bm{\Phi}_{0k} \| +  \| \bm{\Gamma}_k \|)^2 }
		  \ge \| \bm{\Gamma}_k \|_F^2 \left( \beta_{\min}^{-1} + R r_n \right)^{-2}
\end{equation}
where we have used the fact that $\| \bm{\Phi}_{0k} \| = \| \bm{S}_{0k}^{-1} \| = \phi_{\max }(\bm{S}_{0k}^{-1}) = (\phi_{\min }(\bm{S}_{0k}))^{-1} \le \beta_{\min}^{-1} $ and $\| \bm{\Gamma}_k \| \le \| \bm{\Gamma}_k \|_F  \le \| {\bm \Delta} \|_F = R r_n = {\cal O}(r_n)$. 
Therefore,
\begin{equation}
    2 A_1 := \sum_{k=1}^{M_n} A_{1k}  \ge \frac{\sum_{k=1}^{M_n} \| \bm{\Gamma}_k \|_F^2 }
		  { \left(\beta_{\min}^{-1} + R r_n \right)^{2}} =  \frac{\| \bm{\Delta} \|_F^2 }
		  {\left(\beta_{\min}^{-1} + R r_n \right)^{2}}  \label{A1main}
\end{equation}

Turning to $A_{2k}$, we have
\begin{equation}
     |A_{2k}|  \le 2 L_{21k} + 2 L_{22k}
\end{equation}
where
\begin{equation}
      L_{21k} = \big| \sum_{\substack{i,j \\ i \ne j}} [\hat{\bm{S}}_k - \bm{S}_{0k}]_{ij} \Gamma_{kji} \big|
			  = \big| \sum_{\substack{i,j \\ i \ne j}} [\hat{\bm{S}}_k - \bm{S}_{0k}]_{ij}^\ast \Gamma_{kji}^\ast \big| \, ,
\end{equation}
\begin{equation}
      L_{22k} = \big| \sum_{i} [\hat{\bm{S}}_k - \bm{S}_{0k}]_{ii} \Gamma_{kii} \big|
			  = \big| \sum_{i} [\hat{\bm{S}}_k - \bm{S}_{0k}]_{ii}^\ast \Gamma_{kii}^\ast \big| \, .
\end{equation}
To bound $L_{21k}$, using Lemma 3, with probability $> 1- 1/p_n^{\tau-2}$,
\begin{align}
     L_{21k} & \le \| \bm{\Gamma}_k^- \|_1 \, \max_{i,j} \big|  [\hat{\bm{S}}_k - \hat{\bm{S}}_{0k}]_{ij}  \big| 
		 \le  \| \bm{\Gamma}_k^- \|_1 \, C_0 d_{1n}  \, . \label{naeq1205}
\end{align}
Using Cauchy-Schwartz inequality and Lemma 3, with probability $> 1- 1/p_n^{\tau-2}$,
\begin{align}
     L_{22k} & \le \| \bm{\Gamma}_k^+ \|_F \, \sqrt{ \sum_{i=1}^{p_n} \big|  [\hat{\bm{S}}_k - \hat{\bm{S}}_{0k}]_{ii}  \big|^2}   \nonumber \\
		   & \le  \| \bm{\Gamma}_k^+ \|_F \, \sqrt{p_n}  \, 
			              \max_{1 \le i \le p_n} \big|  [\hat{\bm{S}}_k - \hat{\bm{S}}_{0k}]_{ii} \big| \nonumber \\
			 & \le \| \bm{\Gamma}_k^+ \|_F \, \sqrt{p_n}  \, C_0 \, d_{1n}   \nonumber \\
		   & \le    \| \bm{\Gamma}_k^+ \|_F \, C_0 \, d_{2n}  \label{naeq1210}
\end{align}
where $s_{n0}$ is the cardinality of the true edge set ${\cal E}_0$ (see Assumption (A1)). Thus, with probability $> 1- 1/p_n^{\tau-2}$,
\begin{align}
     |A_{2k}| & \le  2 C_0 \big( \| \bm{\Gamma}_k^- \|_1 \,  d_{1n}  
		+  \| \bm{\Gamma}_k^+ \|_F \,  d_{2n} \big)  \, . \label{naeq1200a}
\end{align}
Hence with $A_2 = \sum_{k=1}^{M_n} (1/2) A_{2k}$,
\begin{align}
  |A_2| & \le \sum_{k=1}^{M_n} (1/2) |A_{2k}|  \le  
	  C_0 \, \sum_{k=1}^{M_n} \Big( d_{1n}  \| \bm{\Gamma}_k^- \|_1  
	    +  d_{2n}  \| \bm{\Gamma}_k^+ \|_F \Big) \, . \label{naeq1200}
\end{align}
We now derive an alternative bound on $A_2$. We have w.h.p.\
\begin{align}
  |A_2| & \le  \sum_{i,j=1}^{p_n} \sum_{k=1}^{M_n} \big| [(\hat{\bm{S}} - \bm{S}_{0k}]_{ij} \big| \, 
	           \cdot \big| [{\bm \Gamma}_k]_{ij} \big| \\
	  & \le  C_0 \,  d_{1n} 
		            \sum_{i,j=1}^{p_n} \sum_{k=1}^{M_n} \big| [{\bm \Gamma}_k]_{ij} \big| \\
		& \le  C_0 \,  d_{1n}
		         \sum_{i,j=1}^{p_n} \big(\sqrt{M_n} \| {\bm \Delta}^{(ij)} \|_F \big) \\
		& =  \sqrt{M_n} C_0 \,  d_{1n} 
		   \big( \| \tilde{\bm{\Delta}}^-\|_1 + \|\tilde{\bm{\Delta}}^+\|_1 \big) \label{naeq1201}
\end{align}
where $\tilde{\bm{\Delta}} \in \mathbb{R}^{p_n \times p_n}$ has its $(i,j)$th element $\tilde{{\Delta}}_{ij} = \| {\bm \Delta}^{(ij)} \|_F$.

We now bound $A_{3k}$. Let ${\cal E}_0^c$ denote the complement of ${\cal E}_0$, given by ${\cal E}_0^c = \{ \{i,j\} ~:~ [{\bm S}^{-1}_{0}(f)]_{ij} \equiv 0, ~i\ne j,  ~ 0 \le f \le 0.5 \}$. For an index set ${\bm B}$ and a matrix ${\bm C} \in \mathbb{C}^{p \times p}$, we write ${\bm C}_{\bm B}$ to denote a matrix in $\mathbb{C}^{p \times p}$ such that $[{\bm C}_{\bm B}]_{ij} = C_{ij}$ if $(i,j) \in {\bm B}$, and $[{\bm C}_{\bm B}]_{ij}=0$ if $(i,j) \not\in {\bm B}$. Then $\bm{\Gamma}_k^- = \bm{\Gamma}_{k{\cal E}_0}^- + \bm{\Gamma}_{k{\cal E}_0^c}^-$, and $\| \bm{\Gamma}_k^- \|_1 = \| \bm{\Gamma}_{k{\cal E}_0}^- \|_1 + \| \bm{\Gamma}_{k{\cal E}_0^c}^- \|_1 $. We have
\begin{align}
     A_{3k} & =  \alpha \lambda_n ( \| \bm{\Phi}_{0k}^{-} + \bm{\Gamma}_k^{-} \|_1 - \| \bm{\Phi}_{0k}^{-} \|_1 )  \nonumber \\
		& =  \alpha \lambda_n (  \| \bm{\Phi}_{0k}^{-} + \bm{\Gamma}_{k{\cal E}_0}^- \|_1 
		          + \| \bm{\Gamma}_{k{\cal E}_0^c}^- \|_1
			        - \| \bm{\Phi}_{0k}^{-} \|_1 )  \nonumber \\
			& \ge  \alpha \lambda_n (  \|\bm{\Gamma}_{k{\cal E}_0^c}^- \|_1 - \| \bm{\Gamma}_{k{\cal E}_0}^- \|_1 )  
\end{align}
leading to ($A_3 = \sum_{k=1}^{M_n} A_{3k}$)
\begin{align}
     A_{3} 
			& \ge  \alpha \lambda_n \sum_{k=1}^{M_n} 
			(  \|\bm{\Gamma}_{k{\cal E}_0^c}^- \|_1 - \| \bm{\Gamma}_{k{\cal E}_0}^- \|_1 )  \, .
\end{align}
Similarly, 
\begin{align}
     A_{4} 
			& \ge  (1-\alpha) \lambda_n  
			(  \|\tilde{\bm{\Delta}}_{{\cal E}_0^c}^- \|_1 - \| \tilde{\bm{\Delta}}_{{\cal E}_0}^- \|_1 )  \, .
\end{align}

By Cauchy-Schwartz inequality, $\| \bm{\Gamma}_{k{\cal E}_0}^- \|_1 \le \sqrt{s_{n0}} \| \bm{\Gamma}_{k{\cal E}_0}^- \|_F \le \sqrt{s_{n0}} \| \bm{\Gamma}_{k} \|_F$, hence
\begin{equation}
  \sum_{k=1}^{M_n} \| \bm{\Gamma}_{k{\cal E}_0}^- \|_1 \le \sqrt{M_n s_{n0}} \| {\bm \Delta} \|_F \, .
\end{equation}
Set $\| \bm{\Gamma}_k^- \|_1 = \| \bm{\Gamma}_{k{\cal E}_0}^- \|_1 + \| \bm{\Gamma}_{k{\cal E}_0^c}^- \|_1 $ in $A_2$ of (\ref{naeq1200}) to deduce that w.h.p.\
\begin{align}
  \alpha & A_2 + A_3  \ge -\alpha |A_2| + A_3 \nonumber \\
	 & \ge \alpha(\lambda_n - C_0 d_{1n} ) 
	           \sum_{k=1}^{M_n} \| \bm{\Gamma}_{k{\cal E}_0^c}^- \|_1  \nonumber \\
			& - \alpha ( C_0 d_{1n} + \lambda_n) \sum_{k=1}^{M_n} \| \bm{\Gamma}_{k{\cal E}_0}^- \|_1
			   - \alpha  C_0 d_{2n} \sum_{k=1}^{M_n} \| \bm{\Gamma}_{k}^+ \|_F \nonumber \\
			& \ge - \alpha \Big( ( C_0 d_{1n} + \lambda_n) \sqrt{s_{n0}} 
			   +  C_0 d_{2n} \Big) \sqrt{M_n } \| {\bm \Delta} \|_F 
\end{align} 
where we have used the fact that $\lambda_n \ge  C_0 \sqrt{M_n} \, d_{1n} \ge  C_0 d_{1n}$ and $\sum_{k=1}^{M_n} \| \bm{\Gamma}_{k}^+ \|_F \le \sqrt{M_n } \| {\bm \Delta} \|_F$. Now use $A_2$ of (\ref{naeq1201}) to deduce that w.h.p.\
\begin{align}
  & (1-\alpha)  A_2 + A_4  \ge 
	   (1-\alpha)\Big( (\lambda_n - C_0 \sqrt{M_n} d_{1n} ) \| \tilde{\bm{\Delta}}_{{\cal E}_0^c}^- \|_1  \nonumber \\
			& - ( C_0 \sqrt{M_n} d_{1n} + \lambda_n) \| \tilde{\bm{\Delta}}_{{\cal E}_0}^- \|_1
			   -  C_0 \sqrt{M_n p_n} d_{1n} \| {\bm \Delta} \|_F \Big) \nonumber \\
			& \ge - (1-\alpha) \| {\bm \Delta} \|_F \Big( \lambda_n \sqrt{s_{n0}} +  C_0 \sqrt{M_n}d_{1n} 
			 \big( \sqrt{s_{n0}} +  \sqrt{p_n} \big) \Big) 
\end{align}
where we have used the facts that $\lambda_n \ge  C_0 \sqrt{M_n} d_{1n}$, and $\| \tilde{\bm{\Delta}}_{{\cal E}_0}^- \|_1 \le \sqrt{s_{n0}} \| \tilde{\bm{\Delta}}_{{\cal E}_0}^- \|_F \le \sqrt{s_{n0}} \| {\bm{\Delta}} \|_F$ (by Cauchy-Schwartz inequality). 

Since $r_n = \sqrt{M_n} \, d_{2n} > \sqrt{M_n s_{n0}} \, d_{1n}$, w.h.p.\ we have
\begin{align}
  & A_2 +  A_3 + A_4  \ge - \| {\bm{\Delta}} \|_F \Big( \alpha \big(2 C_0 r_n + \lambda_n \sqrt{M_n s_{n0}} \big) \nonumber \\
	 & \quad  +(1-\alpha)\big(  \lambda_n \sqrt{s_{n0}} + 2 C_0 r_n \big) \Big) \nonumber \\
	& \ge - \| {\bm{\Delta}} \|_F \Big( 2 C_0 r_n + \lambda_n \sqrt{s_{n0}} (\alpha \sqrt{M_n} + (1-\alpha) ) \Big) \nonumber \\
	 & \ge - \| {\bm{\Delta}} \|_F \Big( (2+C_1) C_0 r_n \Big)
	   \label{Armain}
\end{align}
where we have used the fact that, by (\ref{neq15ab1}) and (\ref{neq15abc}), $\lambda_n \sqrt{s_{n0}} (\alpha \sqrt{M_n} + (1-\alpha) ) \le C_1 C_0 r_n $. Using (\ref{Gmain}), (\ref{A1main}) and (\ref{Armain}), and $\| {\bm \Delta} \|_F = R r_n$, we have w.h.p.\
\begin{align}
   G({\bm{\Delta}}) \, \ge & \,  \| \bm{\Delta} \|_F^2 \Big[ \frac{1}{2} \left( \beta_{\min}^{-1} + R r_n \right)^{-2}  
		   - \frac{(2+C_1)C_0}{R}   \Big] \, . \label{naeq1370}
\end{align}
For $n \ge N_2$, if we pick $R$ as specified in (\ref{neq15ab0}), we obtain $R r_n \le R r_{N_2} \le \delta_1/\beta_{\min}$. Then
\begin{align*}
    \frac{1}{ (\beta_{\min}^{-1} + R r_n )^2} & \ge \frac{\beta_{\min}^2}{(1+\delta_1)^2} 
		    = \frac{ 2(2+C_1+ \delta_2)C_0}{R} \\
			&	> \frac{ 2(2+C_1)C_0}{R} \, ,
\end{align*}
implying $G({\bm{\Delta}})  > 0$ w.h.p. This proves (\ref{more10}), hence the desired result (\ref{neq15}).  

Given any $\epsilon >$, pick $\tau > 2$ such that $p_n^{2-\tau} \le \epsilon$ for $n \ge N_4$ for some $N_4$, where $N_4$ exists since $p_n$ is non-decreasing in $n$. Then by (\ref{neq15}), $\| \hat{\bm{\Omega}}_\lambda - \bm{\Omega}_0 \|_F = {\cal O}_P \left( R r_n  \right) 
= {\cal O}_P \left( C_0 C_1 r_n  \right)$. It is easy to see that a sufficient condition for the lower bound in (\ref{neq15abc}) to be less than the upper bound for every $\alpha \in [0,1]$ is  $C_1 = 2(1+\alpha (\sqrt{M_n}-1)) = {\cal O}(\sqrt{M_n})$. By (\ref{naeq58}), $C_0 = {\cal O}(\ln (M_n))$. Therefore, $\| \hat{\bm{\Omega}}_\lambda - \bm{\Omega}_0 \|_F = {\cal O}_P \left( \ln (M_n) \sqrt{M_n} \, r_n  \right)$. This completes the proof of Theorem 1.
$\quad \blacksquare$

\bibliographystyle{unsrt}

\end{document}